\documentclass[prb,aps,twocolumn,eqsecnum,floatfix]{revtex4}
\usepackage{bm}
\usepackage{pstricks}
\usepackage{graphicx}
\usepackage{psfrag}
\usepackage{amsmath,amssymb}
\newlength{\myL}

\newcommand{\beq}{\begin{equation}}
\newcommand{\eeq}{\end{equation}}
\newcommand{\bea}{\begin{eqnarray}}
\newcommand{\eea}{\end{eqnarray}}

\begin{document}

\title{Devil's staircases, quantum dimer models, and stripe formation in strong coupling models of quantum frustration}

\author{Stefanos Papanikolaou}
\author{ Kumar S. Raman}
\author{Eduardo Fradkin}

\affiliation{Department of Physics, University of Illinois at
Urbana-Champaign, 1110 West Green Street, Urbana, Illinois  61801-3080, USA}

\date{\today}

\begin{abstract}
    We construct a two-dimensional microscopic model of interacting quantum 
    dimers that displays an infinite number of periodic striped phases in 
    its $T=0$ phase diagram.  The phases form an incomplete devil's 
    staircase and the %spacing between 
    period becomes arbitrarily large as the 
    staircase is traversed.  The Hamiltonian has purely short-range 
    interactions, does not break 
    any symmetries of the underlying square lattice, and is generic 
    in that it does not involve the fine-tuning of a large number of parameters. Our model, a quantum mechanical analog of the Pokrovsky-Talapov model of fluctuating domain walls in two dimensional classical statistical mechanics,
    %,
    %motivated by the Pokrovsky-Talapov model of fluctuating domain 
   % walls and the 3D anisotropic next nearest neighbor Ising (ANNNI) 
   % model,  
   provides a mechanism by which 
    striped phases with periods large compared to the lattice 
    spacing can, in principle, form in frustrated quantum
    magnetic systems with only short-ranged interactions and no 
    explicitly broken symmetries. 
%    Finally, we discuss how these ideas, obtained for dimer models, may 
 %   be used to construct a local, SU(2)-invariant spin model with the 
  %  same phase diagram.

\end{abstract}

%]

\maketitle

\section{Introduction}

The past two decades have seen the discovery of a number of strongly correlated materials with unconventional physical properties. Due to the competing effects of essentially electronic processes and interactions these doped Mott insulators typically exhibit complex phase diagrams which include antiferromagnetic phases, generally incommensurate charge-ordered phases, and high temperature superconducting phases.  When conducting, these systems do not have well defined electron-like quasiparticles and their metallic states thus cannot be explained by the conventional theory of metals, the Landau theory of the Fermi liquid, and the associated superconducting states cannot be described in terms of the BCS mechanism for superconductivity. 

The startling properties of these materials have led to a number of proposals of non-trivial ground states of strongly correlated systems which share the common feature that they cannot be adiabatically obtained from the physics of non-interacting electrons. A class of proposed ground states are the  resonating valence bond (RVB) spin liquid phases, quantum liquid ground states in which there is no long range spin order of any kind, and the related valence bond crystal phases, of frustrated quantum antiferromagnets\cite{andersonrvb,sr-rvb87} and their descendants.\cite{leenagwen06,senthilfisher1,herm_alg05,mudry94} On the other hand, the presence of competing spatially-inhomogeneous charge-ordered phases in close proximity to both antiferromagnetism and high $T_c$ superconductivity, and the existence of incommensurate low energy fluctuations in the latter phase, strongly suggest that  these phases may have a common origin. It has long been suggested that some form of frustration of the charge degrees of freedom may be at work in these systems.\cite{Emery93,emer99,stripes} The explanation of both the existence of a large pairing scale in the superconducting phase and their close proximity to inhomogeneous 
charge-ordered phases is one of the central conceptual challenges in the physics of these doped Mott insulators.\cite{chapter}

The most studied class of these strongly correlated materials are the cuprate high temperature superconductors (for a recent review on their behavior and open questions see Ref.~\onlinecite{Kivelson03}.) 
Unconventional behaviors have also been seen in other strongly correlated complex oxides.\cite{maeno-mackenzie}  
More recently, strong evidence for non-magnetic phases has been discovered in new frustrated quantum magnetic materials, including the quasi-2D triangular antiferromagnetic insulators 
such as\cite{coldea01}  Cs${}_2$CuCl${}_4$, the quasi-2D triangular organic compounds such 
as\cite{Kanoda03} $\kappa$-(BEDT-TTF)${}_2$Cu${}_2$(CN)${}_3$, and the 3D pyrochlore antiferromagnets such as the spin-ice compound\cite{Ramirez99,Snyder04} Dy${}_2$Ti${}_2$O${}_7$ (although quantum effects do not appear to be prominent in spin-ice systems.) 

It is thus of interest to develop a theoretical framework to describe quantum frustrated systems in the regime of strong correlation, and to understand their role in the mechanism for inhomogeneous phases in strongly correlated systems. This is the main purpose of this paper. It has long been known that generally incommensurate inhomogeneous phases arise in classical systems with competing short range attractive interactions and long range repulsive interactions. In such systems, the short range attractive interactions (whose physical origin depend on the system in question) favor spatially inhomogeneous phases, {\it i.e.\/} phase separation, which is {\em frustrated} by long range (typically Coulomb) repulsive interactions. Coulomb-frustrated phase separation has been proposed as a mechanism for stripe phases in doped Mott insulators\cite{Emery93,low94,lorenzana2001} and in low density electron gases.\cite{jamkivspiv05} Similar ideas were also proposed to explain the structure of the crust of neutron stars, lightly doped with protons,\cite{Ravenhall83,Ravenhall93} and in soft condensed matter ({\it e.g.\/} block copolymers).\cite{seul95}

In this work we will pursue a different approach and consider mechanisms of quantum stabilization ({\it i.e.\/} quantum order by disorder) of stripe-like phases in frustrated quantum systems. We will specifically consider frustrated versions of two-dimensional quantum dimer models,\cite{Rokhsar88} which provide a qualitative description of the physics of quantum frustrated magnets in their spin-disordered phases. The phases that we will discuss here are essentially valence bond crystals with varying degree of commensurability and become asymptotically incommensurate. Since these systems are charge-neutral, there are no long range interactions. As we will see below, quantum fluctuations resolve the high degeneracies of their naive classical limit leading to a non-trivial phase diagram with phases with different degree of commensurability or {\em tilt}. The resulting phase diagram has the structure of an incomplete devil's staircase similar to that found in classical order-by-disorder systems such as the anisotropic next-nearest neighbor Ising (ANNNI) models\cite{fishselke80,bak82}. In the regime in which quantum fluctuations are weak, which is where our calculations are systematically controlled, only a small fraction of the phase diagram exhibits phases with non-trivial modulations. In this ``classical'' regime the observation of non-trivial phases requires fine tuning of the coupling constants. (In contrast, in systems with long range interactions no such fine tuning is needed at the classical level.\cite{low94}) However, as the quantum fluctuations grow, the fraction of the phase diagram  occupied by these non-trivial phases becomes larger. Thus, at finite values of the coupling constants, where our estimates are not accurate, no fine tuning is needed.  

%aqui

%with phase diagrams that are not easily described 
%within the conventional framework of local order parameters and broken 
%symmetries.  This has led to the speculation that some of these 
%materials may display exotic phases in 
%with non-local properties.  
%The most prominent example is the resonating valence bond (RVB) theory 
%of the high $T_{c}$ cuprates\cite{andersonrvb,sr-rvb87} which postulates the existence 
%of 
%spin liquid states (of various sorts).%(more precisely, a long-range RVB phase) within 
%the pseudogap region of the high $T_{c}$ phase diagram.  
%Other exotic
%structures which have been proposed in various experimental or 
%theoretical contexts include the short-range RVB liquid\cite{sr-rvb87}, 
%the algebraic spin 
%liquid\cite{herm_alg05}, phases with fractionalized\cite{senthilfisher1,senthilfisher2} or 
%even nonabelian quasiparticles\cite{nonabelian}, 
%generalized spin liquids based on the projective symmetry 
%group\cite{wen01} and 
%high 
%order striped phases\cite{vojta99,stripes}, which are the focus of this present work.

Considerable progress has been made towards understanding theoretically
the liquid phases, including the proper field 
theoretic description\cite{msf02,ardonne04} which also allows for an analysis of the related valence bond solid phases with varying degrees of commensurability.\cite{vbs04,fhmos04}
%and 
%issues regarding stability\cite{hermU1} and 
%classification\cite{wen01} of the phases and their excitations.
Complementary to this effort is the dynamical question of how, or 
even if, a phase with exotic non-local properties can 
arise in a system where the interactions are purely local.  
We emphasize the requirement of locality because many of these exotic 
structures have been proposed for experimental systems believed to be 
described by a local Hamiltonian (i.\ e.\ of the Hubbard or 
Heisenberg type).  An additional question is whether the exotic 
physics can be realized in an isotropic model or if it is necessary for 
the Hamiltonian to explicitly break symmetries.  
For some of these %exotic 
structures, the dynamical questions have been partially 
settled by the discovery of model Hamiltonians\cite{MStrirvb, 
kitaev05, ardonne04} which stabilize the exotic 
phase over a portion of their quantum phase diagrams.  While many of these 
models do not (currently) have experimental realizations, their value, in 
addition to providing proofs of principle, lies in the 
identification of physical mechanisms, which often have 
validity beyond the specific case considered.  For example, 
the existence of short-range RVB phases was first demonstrated 
analytically in quantum dimer models\cite{Rokhsar88,MStrirvb}, the essential ingredients being geometric 
frustration and ring interactions.  The possibility of such phases 
being realized in spin systems was subsequently demonstrated by the 
construction of a number of spin 
Hamiltonians\cite{herm_pyro04,balfishgirv02,fujimoto05,rms05,damle06,isakov06,sengupta06}, including SU(2) invariant 
models\cite{fujimoto05,rms05}, 
all of which reduce to dimer-like models at low energies. The existence of commensurate valence bond solid phases, with low order commensurability, in doped quantum dimer models\cite{Fradkin90} has been studied recently\cite{Balents05a,Balents05b}, as well as modulated phases in doped quantum dimer models within mean field theory.\cite{vojta99}

\begin{figure}[ht]
{\begin{center}
\includegraphics[width=3in]{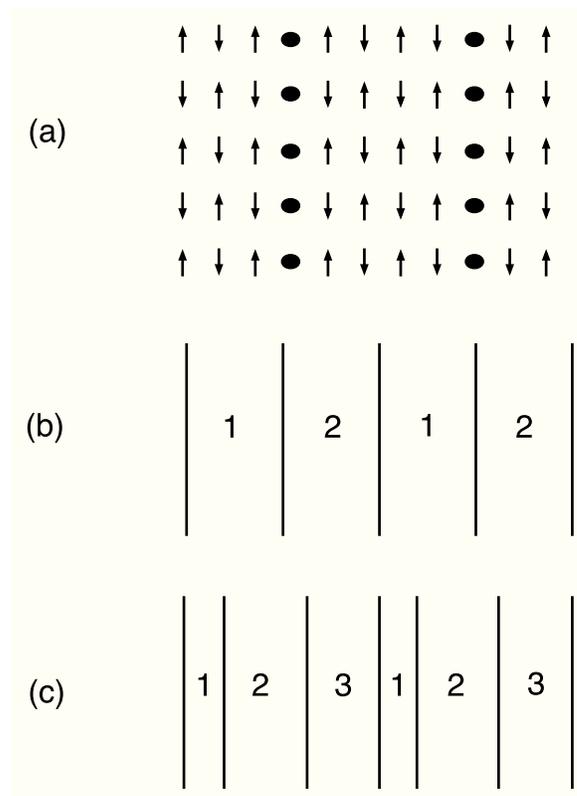}
\caption{Examples of striped phases.  (a) shows stripes of holes 
separating antiferromagnetic domains.  This structure appears in some 
theories of the high $T_{c}$ cuprates.  (b) shows periodically spaced 
domain walls separating regions where the order parameter takes the 
uniform value $\phi_{1}$ or $\phi_{2}$.  (c) is another example where the 
periodicity is associated with a repeating unit cell.  If the repeat 
distance becomes infinite, then the state is said to be {\em 
incommensurate} with the lattice.}
\label{fig:stripe}
\end{center}}
\end{figure}

It is in this context that we ask the following question: is it 
possible to realize high order striped phases in a strongly-correlated quantum system with only 
local interactions and no explicitly broken symmetries?  
While the term {\em stripe} has been used in reference to a number of  
spatially inhomogeous states, here we use the term to 
denote a domain wall between two uniform regions.  The presence of 
domain walls means that rotational symmetry has been broken.  
Fig.~\ref{fig:stripe} gives examples of striped phases.  Theories 
based on the formation of striped phases have been 
proposed in a number of experimental contexts, notably the high 
$T_c$ cuprates\cite{zaanen,zhang,vojta99,stripes}, where the ``stripes'' are lines of doped holes separating antiferromagnetic domains (see Fig.~\ref{fig:stripe}a). In the simplest 
striped states, the domain walls are periodically 
spaced (Fig.~\ref{fig:stripe}b) or are part of a repeating unit cell 
(Fig.~\ref{fig:stripe}c).  We use the term ``high order striped 
phase'' in the case where the periodicity is large compared to
the other characteristic lengths in the model.

Our central result is a positive answer to the question posed in the 
preceding paragraph.  We do this by constructing a two-dimensional 
quantum dimer model, with only short range interactions and no explicitly broken 
symmetries, that shows an infinite number of periodic striped phases in its $T=0$ 
phase diagram.  The collection of states forms an incomplete devil's staircase.  The 
phases are separated by first order transitions and the spacing between stripes 
becomes arbitrarily large as the staircase is traversed.  

Before giving details of the construction, we reiterate that we are 
searching for (high-order) striped phases in a Hamiltonian with only {\em local} 
interactions and {\em without} explicitly breaking any symmetries.  As 
alluded to earlier, a number of experimental systems where 
stripe-based theories have been proposed are widely believed to be 
described by Hamiltonians that meet these restrictions.  A notable 
example is the Emery model \cite{emery87, kivfradgeb04} of the high-$T_{c}$ cuprates, which is a 
generalization of the Hubbard model that includes both Cu and O 
sites.  Since it is not {\em a priori} obvious that a nontrivial 
global ordering such as a high order striped phase will/can unambiguously arise from such 
local, symmetric strongly correlated models, we include these phases in the list of 
``exotic'' structures.  However, in the 
absence of these restrictions, the occurrence of stripe-like 
phases is relatively common.  Relaxing the requirement of high 
periodicity, we note that low order striped quantum phases can occur
in the Bose-Hubbard model at fractional fillings if appropriate 
next-nearest-neighbor interactions are added\cite{sach_compord05}. Relaxing the requirement of 
locality, we note that stripe phases arise naturally in 
systems with long range Coulomb interactions\cite{vojta99,jamkivspiv05}.  More generally, if the 
Hamiltonian includes a term that is effectively a chemical potential 
for domain walls, and if there is a long-range repulsive interaction 
between domain walls, then we may generically expect striped phases where the 
spacing between domain walls is large (compared with the lattice 
spacing).  

A guiding principle in constructing models with exotic phases 
is {\em frustration}, or the inability of a system to simultaneously 
optimize all of its local interactions.  Quantum dimer models are 
relatively simple models that contain the basic physics of quantum
frustration and have proven to be a useful place to look in the search 
for exotic phases\cite{MStrirvb}.  This is one reason why we choose to work in the 
dimer Hilbert space.  A second reason is related to the observation 
that each dimer covering may be assigned a winding number and this 
divides the Hilbert space into topological sectors that are not 
coupled by local dynamics.  The ground state wavefunction of a dimer 
Hamiltonian will typically live in one of these sectors (ignoring the 
complications of degeneracy for the moment).  As 
parameters in the Hamiltonian are varied, it is possible that at some 
critical value, the 
sector containing the ground state will change.  Such a scenario, 
which occurs even in the simplest dimer model formed by Rokhsar and Kivelson, is 
an example of a {\em quantum tilting transition} between a ``flat'' 
and ``tilted'' state.  In 
Refs.~\onlinecite{vbs04} and \onlinecite{fhmos04}, it was shown, by 
field theoretic arguments, that 
when such a transition is perturbed, states of ``intermediate 
tilt'' (this will be made more precise below) may be stabilized.  As 
we will show, these intermediate tilt states may be viewed as 
stripe-like states, of the form we are interested in.  Taking 
inspiration from these ideas, our construction involves perturbing 
about a tilting transition in a specially constructed dimer model.  

 In classical Ising systems, the competition between nearest-neighbor and 
next-nearest-neighbor interactions is a 
well-known mechanism for generating incommensurate phases
\cite{fishselke80}. 
Classical models of striped phases are 
based on two complementary principles: soliton formation and competing 
interactions\cite{bak82}.  Our quantum construction is based on 
analogies with two classical models that are representatives of these 
two aspects: the Pokrovsky-Talapov model of fluctuating domain 
walls\cite{poktal78} and the ANNNI model\cite{fishselke80}.  In section 
\ref{sec:classical}, we review 
the relevant features of these models. In 
section \ref{sec:dimer}, we review the salient features about dimer 
models and tilting transitions. In section \ref{sec:overview}, we give an 
overview of the construction and technical details are presented in 
section \ref{sec:details} and the appendices.  In section 
\ref{sec:spin}, we discuss how these ideas connect to spin models, thus
extending our ``proof of principle'' to systems with physical degrees of 
freedom.  We discuss implications of these results 
in section \ref{sec:discuss}.  In two appendices we give technical details of our calculations.

\section{Classical models}
\label{sec:classical}

Our approach builds on principles underlying 
stripe formation in classical models, where the problem is also 
referred to as a commensurate-incommensurate transition\cite{bak82}.
The classical models are based on two complementary principles: the 
insertion of domain walls and competing local interactions.  

\begin{figure}[t]
{\begin{center}
\includegraphics[width=3in]{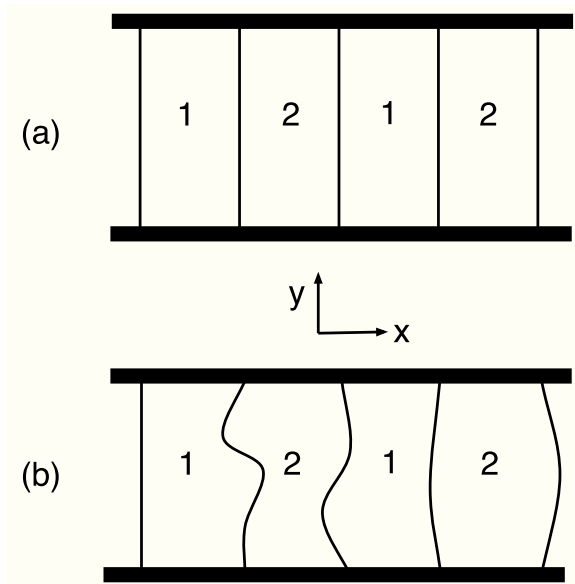}
\caption{(a) ground and (b) excited states of the
Pokrovsky-Talapov model of fluctuating classical domain walls.  
This is a two-dimensional anisotropic model where domain walls form 
along the $y$ direction and separate regions where the order parameter 
is uniform.  While the domain walls cost energy, they 
are allowed to fluctuate, which carries entropy.  For a range of 
parameters, the domain walls are actually favored by the free energy 
minimization.}
\label{fig:poktal}
\end{center}}
\end{figure}
\begin{figure}[h!]
{\begin{center}
\includegraphics[width=2in]{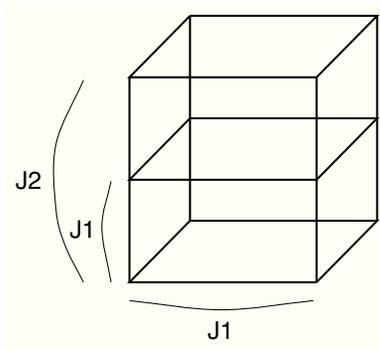}
\caption{The anisotropic next nearest neighbor Ising (ANNNI) model.  
Ising spins lie on the points of a $d$-dimensional cube.  Nearest 
neighbor interactions are ferromagnetic ($J_{1}<0$.  Along one of the 
directions, we also have antiferromagnetic next nearest neighbor 
interactions $J_{2}>0$.}
\label{fig:annni1}
\end{center}}
\end{figure}

A toy model relevant to the present work is the picture of fluctuating 
domain walls in two dimensions, introduced by Pokrovsky and 
Talapov\cite{poktal78}.  The walls are allowed to fluctuate though the 
ends are fixed (Fig.~\ref{fig:poktal}), which precludes bubbles.  The 
free energy minimization is a competition between the creation energy 
of having walls, the elastic energy of deviating from the flat state, and 
the entropic benefit of allowing the walls to fluctuate.   This theory predicts 
a transition from a uniform phase to a striped phase.  The spacing between 
walls depends on the parameters of the theory (including 
temperature) and can be large compared with other length scales.     

A second model relevant to the present work is the classical anisotropic 
next-nearest-neighbor Ising (ANNNI) model in three (and higher) 
dimensions\cite{fishselke80,bakvonboehm80}.  This model 
(Fig.~\ref{fig:annni1}) describes 
Ising spins on a cubic lattice with ferromagnetic nearest-neighbor 
interactions $J_{1}<0$ and antiferromagnetic next-nearest-neighbor 
interactions $J_{2}>0$ {\em along one of the lattice directions}.  
A key feature of 
the ANNNI Hamiltonian is a special 
point $J_{1}/J_{2}=2$ where a large number of stripe-like states are 
degenerate at zero temperature.  As the temperature is raised, the 
competition between $J_{1}$ and $J_{2}$ causes an infinite number of
modulated phases to emerge from this degenerate point.  
The phase diagram in the low $T$ limit was studied analytically in 
Ref.~\onlinecite{fishselke80} using a novel perturbative scheme where the 
existence of higher order phases was established at successively high orders 
in the perturbation theory.  Numerical studies at higher 
temperatures\cite{bakvonboehm80} indicated that incommensurate phases 
occur near the phase boundaries.  Therefore, the collection of phases form an 
incomplete devil's staircase\cite{bak82,3dannni}.  A quantum version of the ANNNI
model was studied in Refs.~\onlinecite{yeomans95} and \onlinecite{harris95}.  

The phase diagram of our model is similar to that of the ANNNI model 
and our analytical methods are similar in spirit to 
Ref.~\onlinecite{fishselke80}.  
However, the basic physics of our model corresponds more clearly to a quantum 
version of the energy-entropy balance occuring in the fluctuating domain wall 
picture.  We now discuss one more ingredient of the construction 
before putting the pieces together in section \ref{sec:overview}.  

\section{Quantum dimer models and tilting transitions}
\label{sec:dimer}

A hard-core dimer covering of a lattice is a mapping where each site 
of the lattice forms a bond with exactly one of its nearest 
neighbors.  Each dimer covering is a basis vector 
in a dimer Hilbert space and the inner product is such that 
different dimer coverings are orthogonal.  Quantum dimer models are 
defined on this dimer Hilbert space through operators that manipulate 
these dimers in ways that preserve the hard-core condition.  These 
models were first proposed as effective descriptions of the strong coupling 
regime of quantum spin systems\cite{Rokhsar88} and 
Refs.~\onlinecite{balfishgirv02,fujimoto05,rms05}
discuss ways in which this correspondence can be made precise.

\begin{figure}[ht]
{\begin{center}
\includegraphics[width=3in]{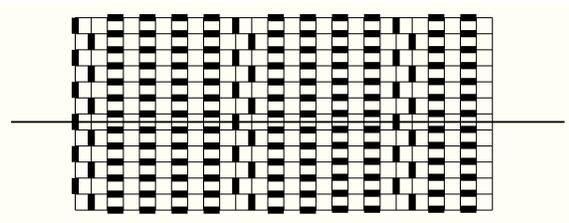}
\caption{Winding numbers: draw a reference line that extends through and 
around (due to the periodic boundary condition) the system and 
label the vertical lines of the lattice alternately as $A$ and $B$ 
lines.  For any dimer configuration, we may define, with regard to 
this reference line, the winding 
number $W_{x}=N_{A}-N_{B}$, where $N_{A}$ is the number 
of $A$ dimers intersecting the line and similarly for $N_{B}$.
We can similarly draw a vertical line and define a similar quantity $W_{y}$.
Note that this particular construction works for a 2d bipartite lattice.  
For 2d non-bipartite lattices, the construction is simpler: count the 
total number of dimers intersecting the horizontal and vertical reference lines 
and there are four sectors corresponding to whether $W_{x,y}$ is even 
or odd.}
\label{fig:tilt}
\end{center}}
\end{figure}

The space of dimer coverings can be subdivided into different 
topological sectors labelled by the pair of winding numbers 
$(W_{x},W_{y})$ defined in Fig.~\ref{fig:tilt}.  The winding number 
is a global property in that it is not affected by any local 
rearrangement of dimers.  In particular, for any local Hamiltonian, 
the matrix element between dimer coverings in different sectors will be zero.

\begin{figure}[ht]
{\begin{center}
\includegraphics[width=3in]{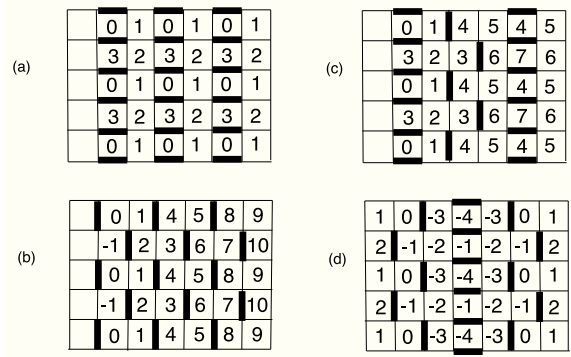}
\caption{Sample dimer configurations with corresponding height 
mappings.  The height mapping involves assigning integers to the squares of the 
lattice in the following manner.  Divide the bipartite lattice into $A$ and $B$ 
sublattices.  Assign zero height to a reference square and then 
moving clockwise around an $A$ site, the height increases by one if a 
dimer is not crossed and decreases by three if a dimer is crossed.  
The same rule applies moving counterclockwise about a $B$ site.  The 
integers correspond to local heights of a crystal whose base lies on 
the page.  In these examples, the lower square in the second column is taken as 
the reference square.  (a) Dimers are arranged in 
columns corresponding to a surface that is flat on average (though 
there are fluctuations at the lattice scale).  (b) Dimers are 
staggered and the corresponding surface is maximally tilted.  (c) 
There average tilt is nonzero due to the staggered strip in between 
the flat columnar regions.  (d)  Going from left to right, the surface 
initially falls and then rises giving an average tilt of zero.  This 
is because the two staggered regions have opposite orientation.}
\label{fig:height}
\end{center}}
\end{figure}

For a given local Hamiltonian, the ground state of the system will 
typically lie in one of the topological sectors.  A common occurrence in dimer 
models is a quantum phase transition in which the topological sector 
containing the ground state changes.  Such an occurrence is 
called a tilting transition because a dimer covering of a 2d bipartite 
lattice may be viewed as the surface of a three-dimensional crystal 
through the height representation\cite{blote82}.  In this language, the 
different topological sectors correspond to different values for 
the (global) average tilt of the surface.  The correspondence between 
dimers and heights is reviewed in Fig.~\ref{fig:height} but for the 
present purpose, it is sufficient to {\em define} the ``tilt'' of a dimer 
covering as its ``winding number per unit length''.  The simplest dimer model 
introduced in Ref.~\onlinecite{Rokhsar88}, has a tilting transition between a 
flat state (zero tilt) and the staggered state, which is maximally tilted 
(Fig.~\ref{fig:height}b).  At the critical point, called the 
Rokhsar-Kivelson or RK point, the Hamiltonian has a ground state 
degeneracy where all tilts are equally favored. 

The recognition of the Rokhsar-Kivelson dimer model as a tilting 
transition led to a field theoretic description \cite{hen97,msf02} 
of the RK point based on a coarse-grained\cite{action} version of the height field 
(Fig.~\ref{fig:height}).  The stability of this field theory 
was studied in Refs.~\onlinecite{vbs04} and \onlinecite{fhmos04}.  These studies 
showed that by tuning a small perturbation and non-perturbatively adding irrelevant operators, it is 
possible to make the tilt vary continuously from a
flat state to the maximally tilted state.  In addition, it was 
observed that the system has a tendency to ``lock-in'' at values of the 
tilt commensurate with the underlying microscopic lattice, 
the specific values depending on details of the perturbation.  It was 
also noted that while a generic perturbation would make the transition first 
order\cite{vbs04}, for a sufficiently small perturbation, the correlation length 
was extremely large\cite{fhmos04} which, it was argued, justified the 
field theory approach nonetheless.  Therefore, the generic effect of 
perturbations would be to smoothen the Rokhsar-Kivelson 
tilting transition by making the system pass through an incomplete devil's 
staircase of intermediately tilted states.  One may suspect that the structure 
of the field theory, including the predictions of 
Refs.~\onlinecite{vbs04} and \onlinecite{fhmos04}, would hold for a broader class of tilting 
transitions.  In particular, one may consider 
the case where the critical point is merely a point of large 
degeneracy where all tilts are favored\cite{tilt, nussinov}, which is analogous 
to the classical ANNNI model.  

The relevance of all of this to the present work is most easily seen 
by Fig.~\ref{fig:ideal}, which shows the simplest examples of states 
that have intermediate tilt (i.\ e.\ winding number).  These are 
stripe-like states having a finite density of staggered domain walls 
and more general tilted states may be obtained by locally rearranging 
the dimers.  The preceding discussion suggests that these kinds of 
structures arise naturally when quantum dimer models are perturbed 
around a tilting transition.  This observation will guide the 
construction outlined in the next section.

\section{Overview of strategy} 
\label{sec:overview}

We now combine various ideas presented in sections \ref{sec:classical} and 
\ref{sec:dimer} to construct the promised quantum model.  In the 
present section, we present an overview of the construction with 
details and subtleties relegated to section \ref{sec:details}.  The basic idea is to 
construct a quantum dimer model with a tilting transition and then to 
appropriately perturb this model to realize a staircase of striped 
states. We design the unperturbed system to have a large degeneracy at the 
critical point, with each of the degenerate states having
a domain-wall structure.  The perturbation will effectively make 
these domain walls fluctuate and the degeneracy will be lifted in a 
quantum analog of the energy-entropy competition that drives the 
classical Pokrovsky-Talapov transition.  Using standard quantum 
mechanical perturbation theory, we will 
obtain a phase diagram similar to the classical 3D ANNNI model and 
will find that phases with increasingly long periodicities
will be stabilized at higher orders in the perturbation theory.  This 
mathematical approach is similar in spirit to the analysis of the 
classical ANNNI model in Ref.~\onlinecite{fishselke80}.   

A simple, rotationally invariant Hamiltonian with a tilting transition is given by:
\begin{equation}
\scalebox{0.43}{\includegraphics{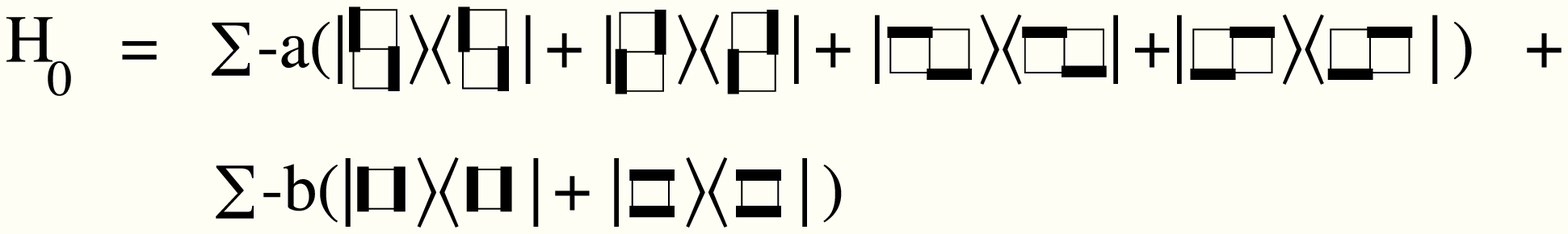}}\\
\label{eq:parent4}
\end{equation}
This model displays a first-order transition between a columnar and fully staggered state at a very degenerate point, where $2a=b$.  In principle, we may perturb this model
with an off-diagonal resonance term and expect phases with intermediate tilt (and 
possibly other exotic phases) on the general grounds discussed in the previous 
section.  However, it is difficult to make precise statements about 
the phase diagram even for such fairly simple models.  We will study a slightly
constrained version of this model that is convenient for making analytical 
progress.      

\begin{figure}[ht]
{\begin{center}
\includegraphics[width=3in]{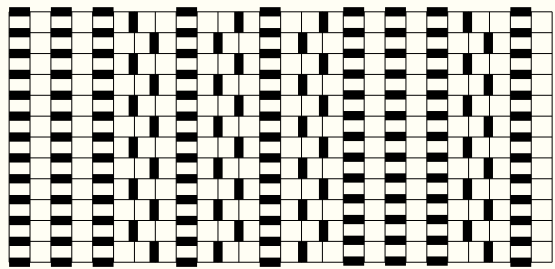}
\caption{A typical domain wall state selected by $H_{0}$.  
These states break translational and rotational symmetry.  The staggered 
strips, which are one column wide and may have one of two orientations, are like 
domain walls separating columnar regions, which may have arbitrary width.
When $a=b$ in Eq.~ \ref{eq:energy0}, the set of these states spans the 
degenerate ground state manifold of $H_{0}$.}
\label{fig:generic}
\end{center}}
\end{figure}

We construct the quantum dimer model in two steps.  First, we 
construct a diagonal parent Hamiltonian $H_{0}$ (Eq.~\ref{eq:H0}) where the 
ground states 
are separated from excited states by a tunably large gap.   $H_{0}$ will 
not break any lattice symmetries, but the preferred ground states will 
spontaneously break translational and rotational symmetry.
In particular, we design $H_{0}$ to select ground states having the
domain wall structure shown in Fig.~\ref{fig:generic}. 
In these states, the dimers arrange themselves into staggered domains of unit 
width separating columnar regions of arbitrary width.  The columnar dimers are 
horizontal if the staggered dimers are vertical (and vice versa) and the 
staggered strips come in two orientations.  Notice that the fully columnar 
state (a columnar region of infinite width) is included this collection but 
the fully staggered state (Fig.~\ref{fig:height}b), which appears in the 
Rokhsar-Kivelson phase diagram, is not.  In the following, we will 
typically draw the staggered strips as vertical domain walls but the 
horizontal configurations are equally possible.   

\begin{figure}[ht]
{\begin{center}
\includegraphics[width=3in]{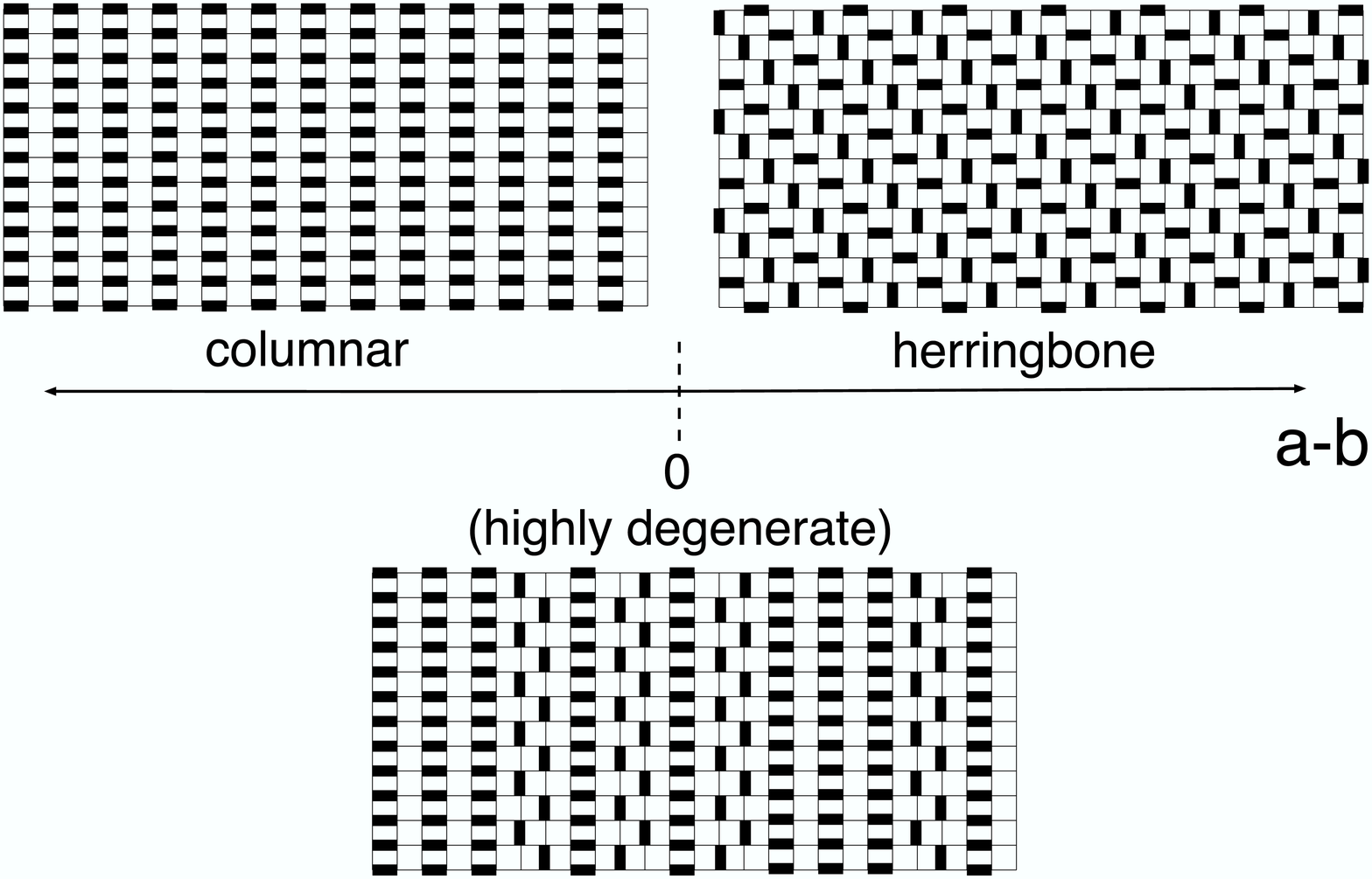}
\caption{Ground state phase diagram of the parent Hamiltonian $H_{0}$ as a 
function of the parameter $a-b$ (see Eq.~\ref{eq:H0}).  In these 
states, the dimers may only have two attractive bonds.  When $a-b=0$, the 
states of Fig.~\ref{fig:generic} are degenerate ground states.  Away from this 
point, the system enters a state where dimers either maximize or 
minimize the number of staggered interactions.  The maximally staggered 
configuration is commonly called the ``herringbone'' state.}
\label{fig:phase0}
\end{center}}
\end{figure}

In analogy with the ANNNI model, $H_{0}$ is designed so that all of 
these domain wall states are degenerate when the couplings are tuned 
to a special point.  Away from this point, the system will enter either a flat 
or a tilted phase.  The unperturbed 
phase diagram is sketched in Fig.~\ref{fig:phase0}.    

\begin{figure}[ht]
{\begin{center}
\includegraphics[width=3in]{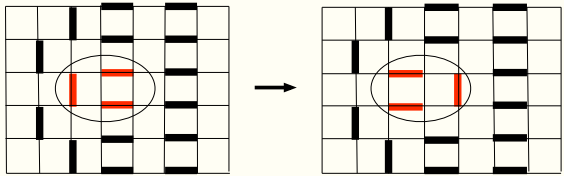}
\caption{One of the terms in the operator of Eq.~\ref{eq:V} will flip 
the circled cluster as shown.}
\label{fig:flip}
\end{center}}
\end{figure}

The second step of the construction is to perturbatively add a small, 
non-diagonal, resonance term $tV$:
\begin{equation}
\psfrag{t}{$t$}
\scalebox{0.5}{\includegraphics{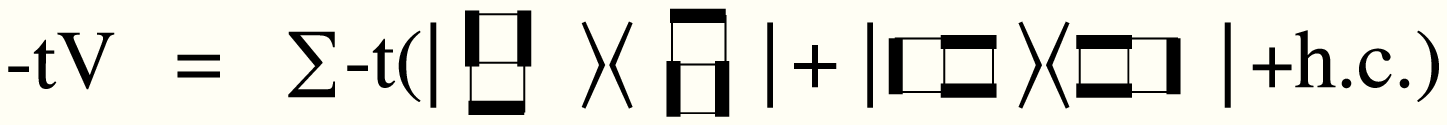}}\\
\label{eq:V}
\end{equation}
The sum is over all plaquettes in the lattice.  Depending on the  
local dimer configuration of the wavefunction, the individual terms in 
this sum will either annihilate the state or flip the local cluster of 
dimers as shown in Fig.~\ref{fig:flip}.  The action of this operator 
on the domain wall states (Fig.~\ref{fig:generic}) is confined to the 
boundaries between staggered and columnar regions and effectively makes 
the domain walls fluctuate.  Notice that Eq.~\ref{eq:V} 
is equivalent to two actions of the familiar Rokhsar-Kivelson two-dimer 
resonance term.  We expect the basic conceptual argument to apply for a more 
general class of perturbations, including the two-dimer resonance, but we consider 
the specific form of Eq.~\ref{eq:V} to simplify certain technical aspects of the calculation.
We will elaborate on this more in the next section.

\begin{figure}
    \begin{tabular}{cc}
	\centering
	\begin{minipage}{1.5in}
	    \includegraphics[width=1.5in]{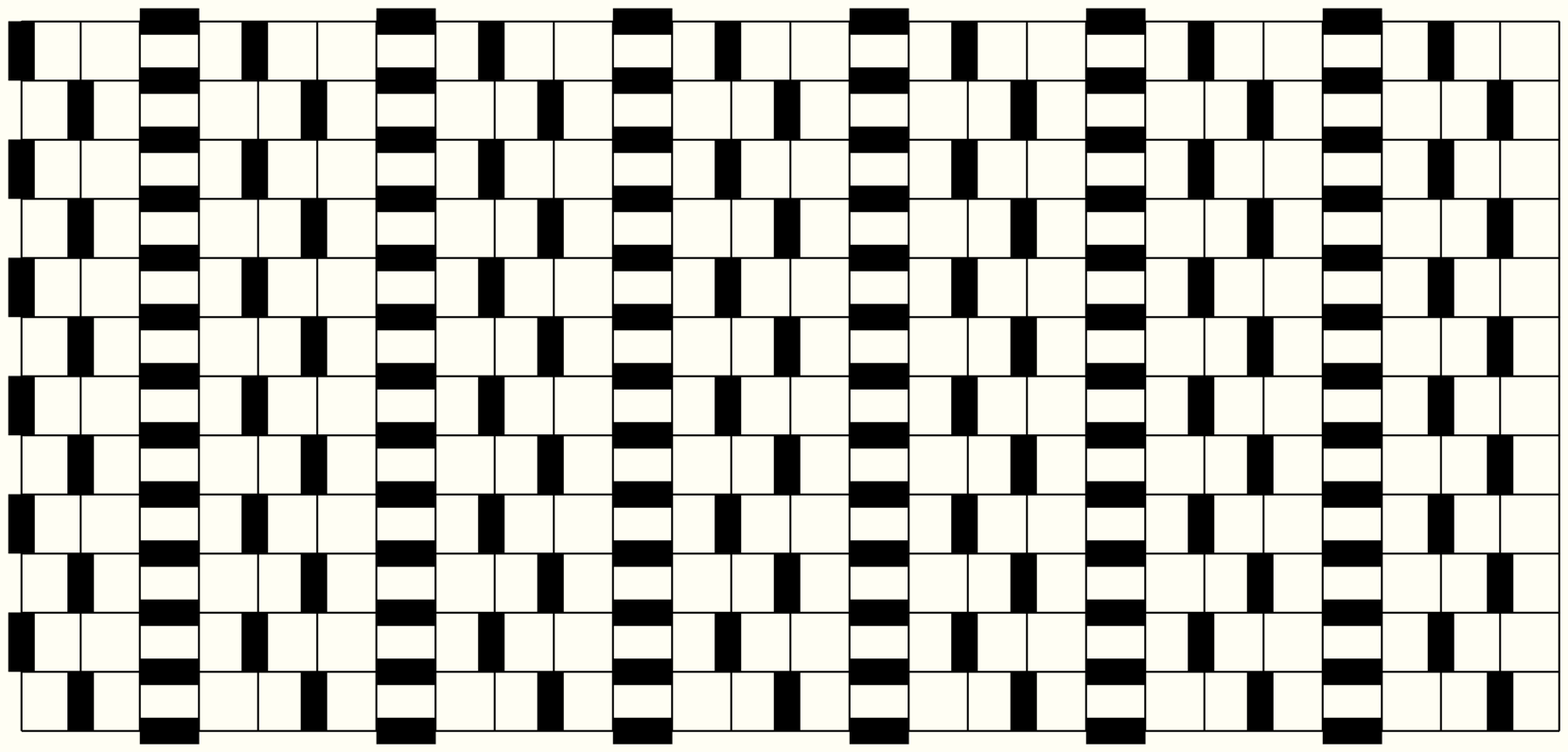}
	 \end{minipage}&
	 \begin{minipage}{1.5in}
	    \includegraphics[width=1.5in]{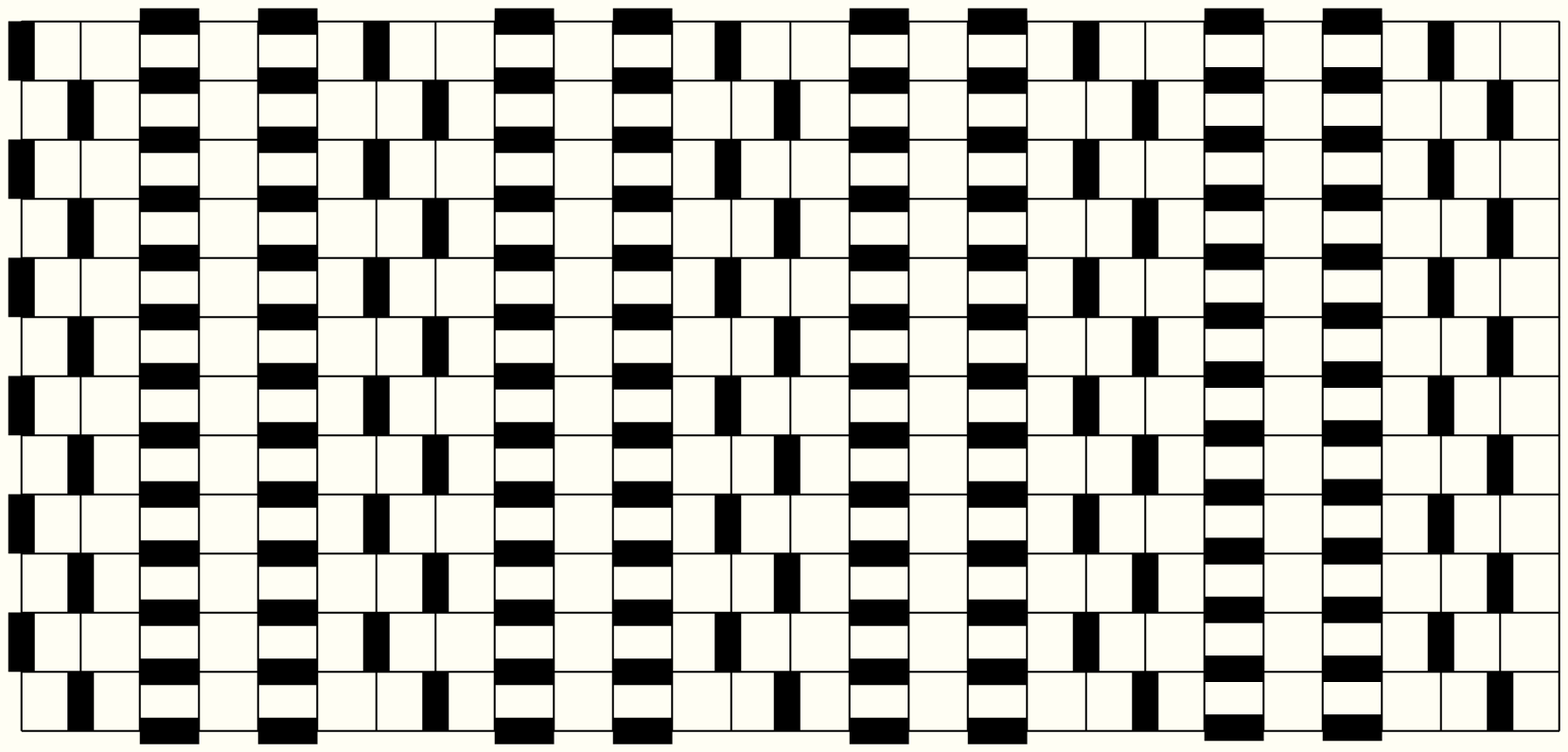}
	 \end{minipage}\\ (a) & (b) \\
	 \begin{minipage}{1.5in}
	     \includegraphics[width=1.5in]{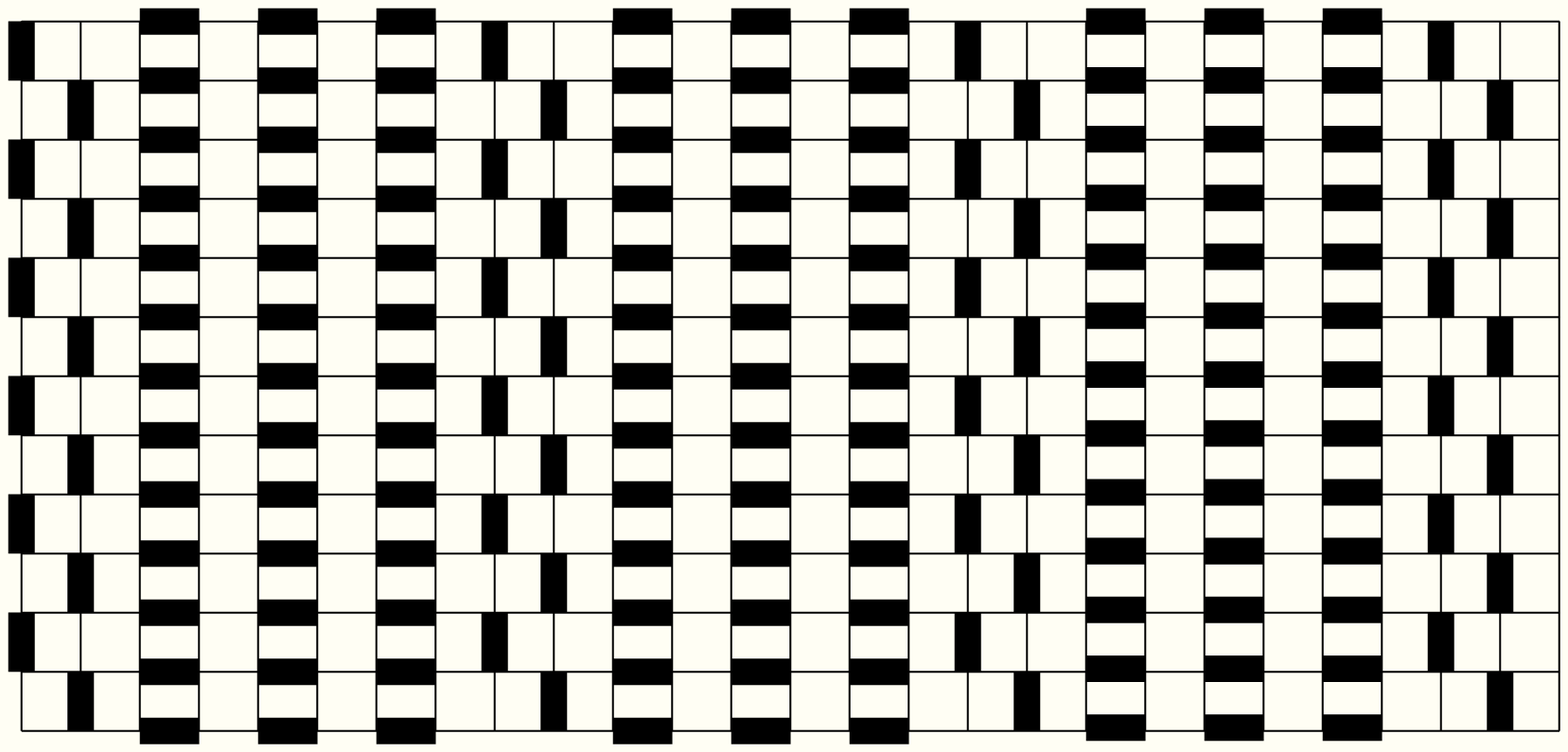}
	  \end{minipage}&
	  \begin{minipage}{1.5in}
	     \includegraphics[width=1.5in]{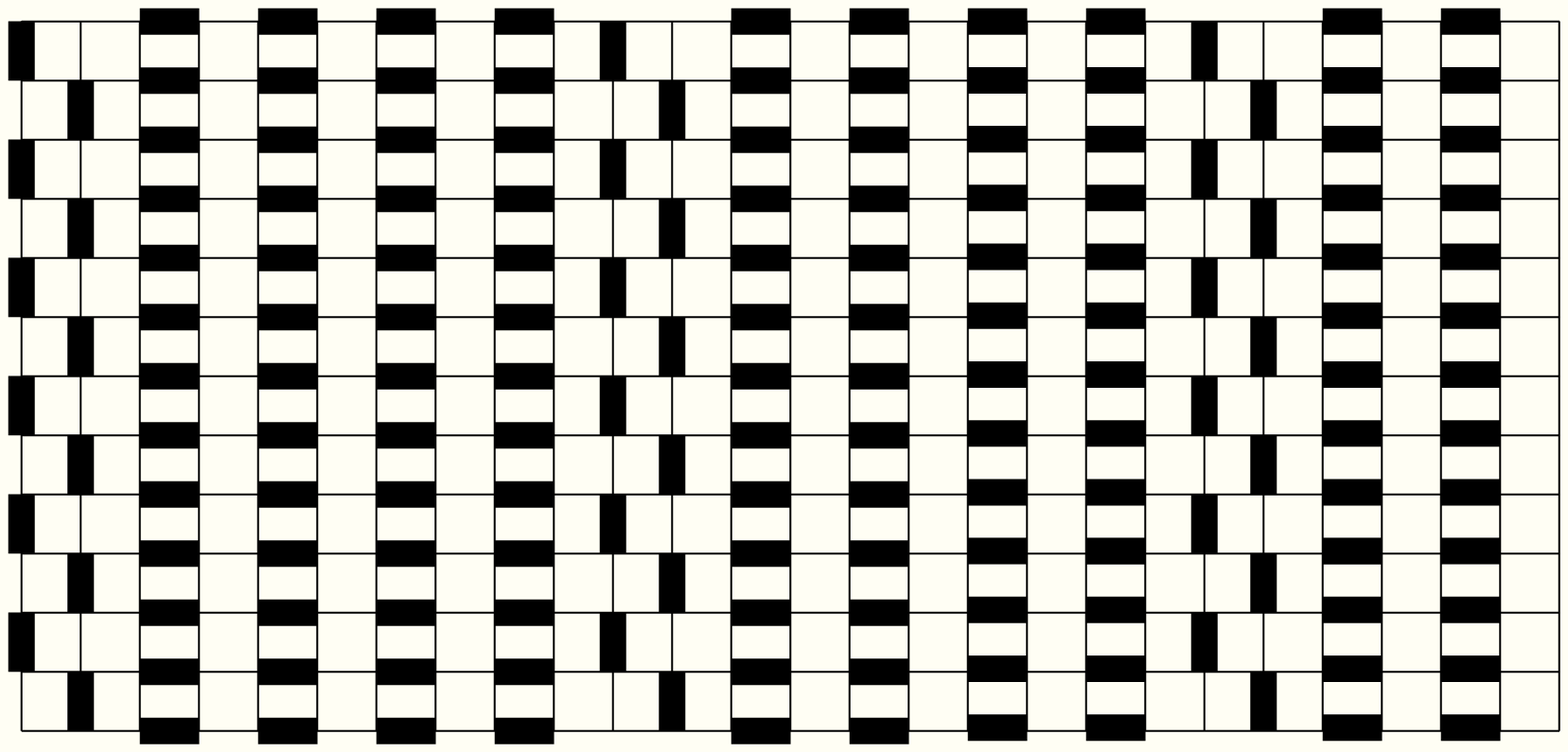}
	  \end{minipage}\\ (c) & (d) \\
	  \begin{minipage}{1.5in}
	      \includegraphics[width=1.5in]{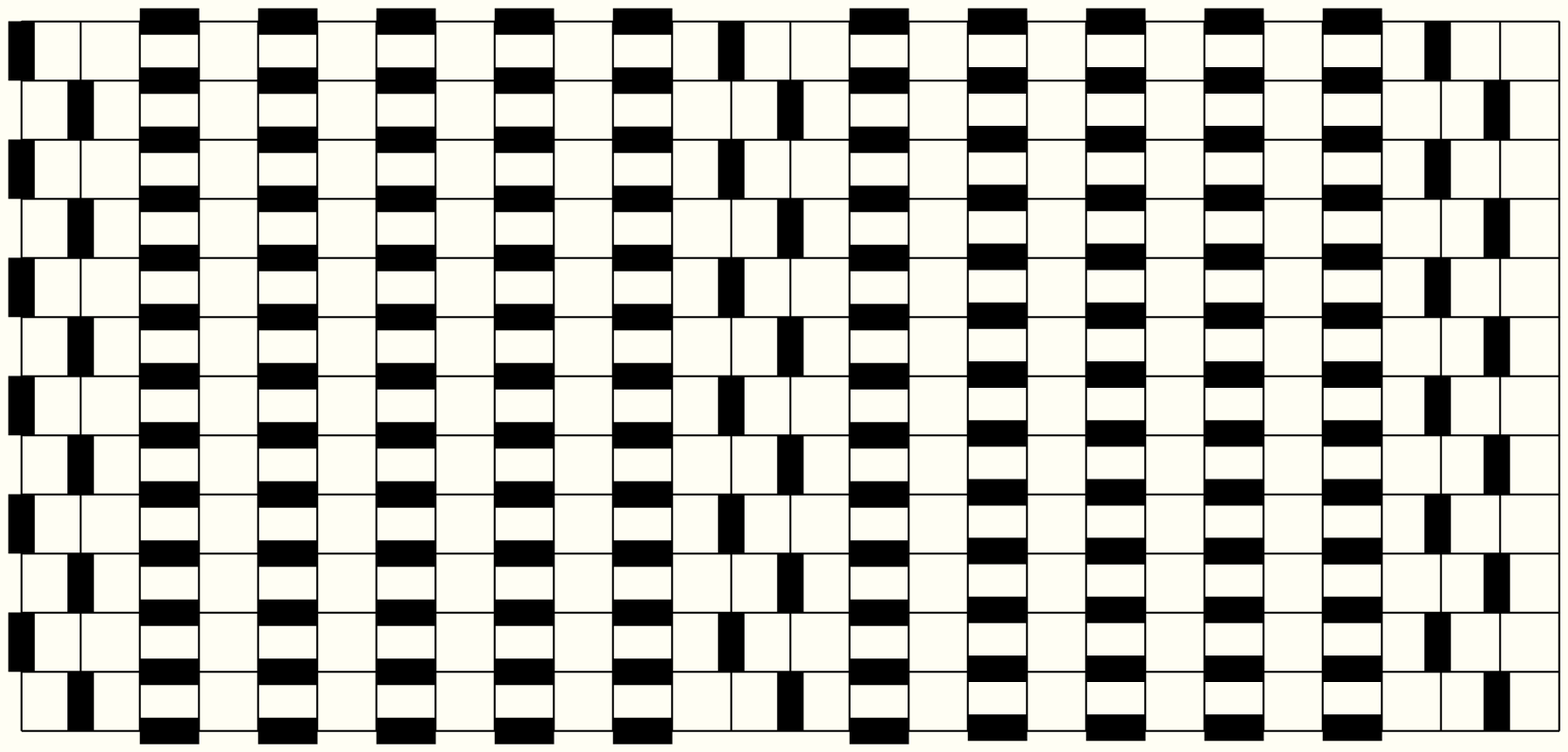}
	   \end{minipage}&
	   \begin{minipage}{1.5in}
	      \includegraphics[width=1.5in]{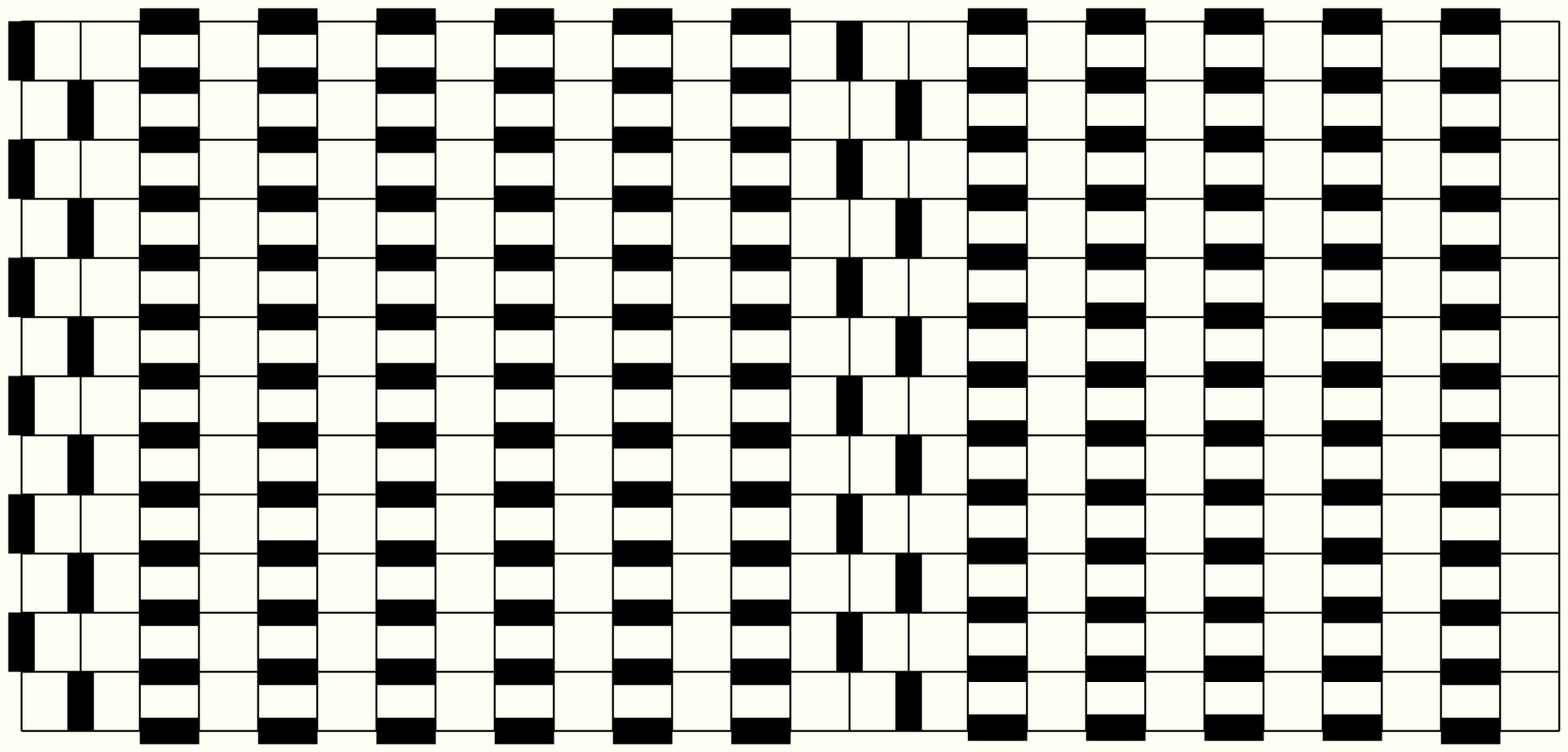}
	   \end{minipage}\\ (e) & (f) \\	 
    \end{tabular}	 
\caption{Examples of ideal tilted states.  In these states, the 
domain walls have the same orientations and are uniformly spaced.  The 
notation 
[1n] denotes the state where one staggered strip is followed by 
$n$ columnar strips and so on.  It is understood that [1n] 
collectively refers to the above states and those related by translational,
rotational, and reflection (i.e.\ switching the orientation of 
the staggering) symmetries.  The examples drawn here, where it is understood 
that what we are seeing is part of a larger lattice, are (a) [11] (b) 
[12] (c) [13] (d) [14] (e) [15] (f) [16].}
\label{fig:ideal}
\end{figure}

\begin{figure}[ht]
%\vspace{-2.5cm}
{\begin{center}
\includegraphics[width=0.37\textwidth]{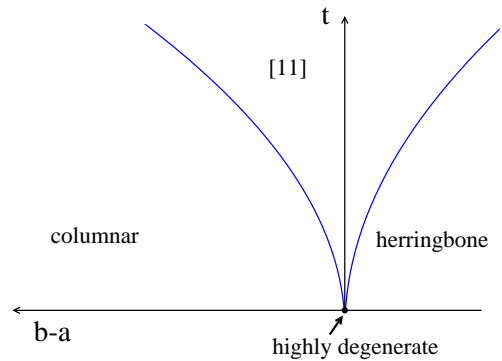}
%\includegraphics[width=3in]{stairs2.eps}
%%\vspace{-1cm}
\caption{(Color online) Ground state phase diagram of $H=H_{0}+tV$ from second order 
perturbation theory.}
\label{fig:phase2}
\end{center}}
\end{figure}

The degenerate point of Fig.~\ref{fig:phase0} may be viewed as the 
degeneracy of an individual vertical column having a staggered or 
columnar dimer arrangement.  The perturbation \ref{eq:V} lifts this 
degeneracy by lowering the energy of configurations with 
domain walls by an amount of order $\sim Lt^{2}$ per domain wall, where 
$L$ is the linear size of the system.  Therefore, the system favors 
one of the states with a maximal number of domain walls and for technical 
reasons discussed below, will choose 
the one having maximal tilt: the [11] state in Fig.~\ref{fig:ideal}a.   
However, the degeneracy between columnar and staggered strips will be 
restored by detuning $H_{0}$ from the $t=0$ degenerate point.  This 
implies the second order phase diagram sketched in Fig.~\ref{fig:phase2}.

\begin{figure}[ht]
{\begin{center}
\includegraphics[width=3in]{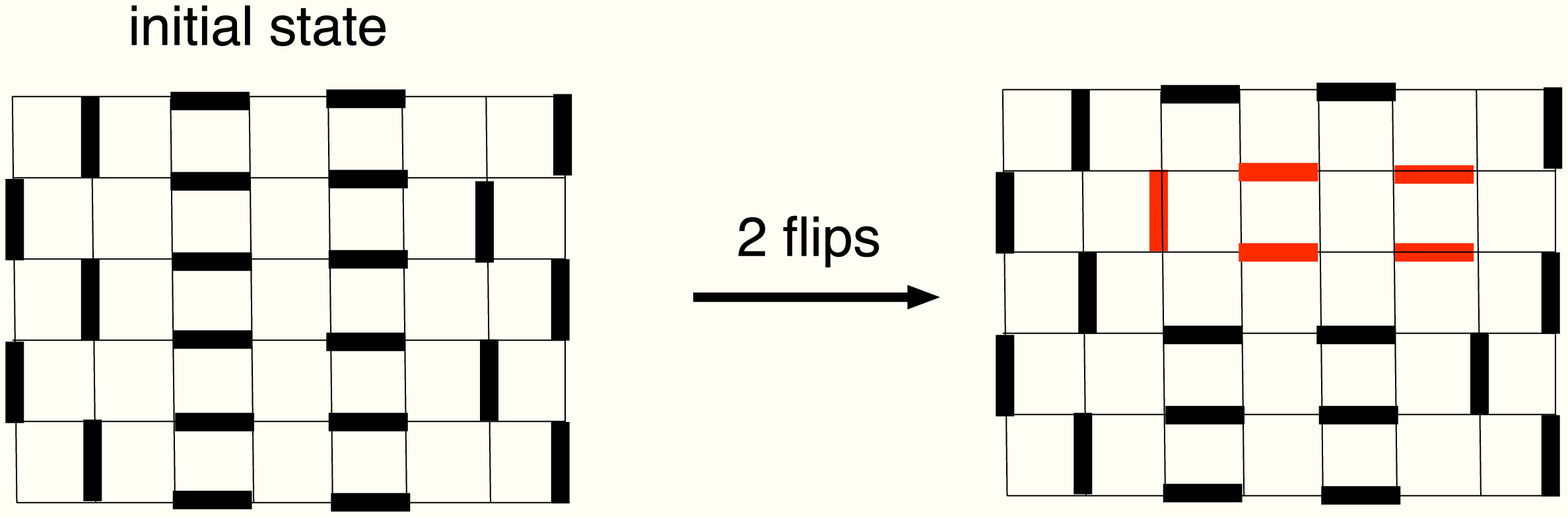}
\caption{(Color online) The excited state on the right is obtained from the initial 
state by acting twice with the perturbation in Eq.~\ref{eq:V}.}
\label{fig:four}
\end{center}}
\end{figure}

\begin{figure}[h!]
%%\vspace{-2.5cm}
{\begin{center}
\includegraphics[width=0.37\textwidth]{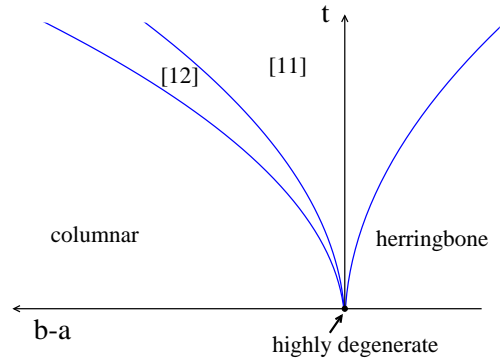}
%\includegraphics[width=3in]{stairs4C.eps}
%%\vspace{-1cm}
\caption{(Color online) Ground state phase diagram of $H=H_{0}+tV$ from fourth order 
perturbation theory.  The width of the [12] phase is order $t^{4}$.}
\label{fig:phase4}
\end{center}}
\end{figure}

Fig.~\ref{fig:phase2} is correct up to error terms of order $t^{4}$.  
To this approximation, states having tilt in between the [11] state 
and the (flat) columnar state are degenerate on the phase boundary.  
Physically, this boundary occurs when the energy from second order 
processes which stabilize the staggered domains is precisely balanced by the
energy of a columnar strip.  This degeneracy will be partially lifted by going 
to higher orders in perturbation theory.  We find that at fourth order, a new 
phase is stabilized in a region of width $\sim t^{4}$ between the columnar and 
[11] phases.  This new phase is the one which maximizes the number of 
fourth order resonances shown in Fig.~\ref{fig:four} and is the [12] 
state (Fig.~\ref{fig:ideal}b).  The corrected phase diagram is given 
in Fig.~\ref{fig:phase4}.     

Fig.~\ref{fig:phase4} is accurate up to corrections of order $t^6$.  
To this approximation, on the boundaries, states with tilts in between the
bordering phases are degenerate.  These degeneracies will be partially 
lifted at higher orders in the perturbation.  At higher orders, there will be 
resonances corresponding to increasingly complicated fluctuations of the 
staggered lines but at $n$th order, the competition between the [1,n-1] 
and columnar phases will always be decided by the $n$th order 
generalization of the long resonance in Fig.~\ref{fig:four}.
The competition will stabilize a new [1n] phase 
in a tiny region of width $\sim t^{2n}$ between the [1,n-1] and columnar 
phases resulting in the phase diagram of Fig.~\ref{fig:phaseN}.

\begin{figure}[t]
%%\vspace{-2cm}
{\begin{center}
\includegraphics[width=0.37\textwidth]{stairs_phaseN-v2.eps}
\%%%vspace{-1cm}
\caption{(Color online) Ground state phase diagram of $H=H_{0}+tV$ from nth order 
perturbation theory.  The width of the [1n] phase is of order $t^{2n}$.}
\label{fig:phaseN}
\end{center}}
\end{figure}
\begin{figure}[t]
%%\vspace{-1.7cm}
{\begin{center}
\includegraphics[width=0.37\textwidth]{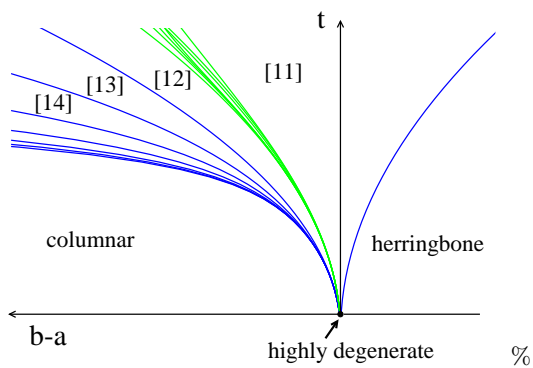}
\%%%{-1.3cm}
\caption{(Color online) The boundaries of the [1n] sequence will typically open into
finer phases and subsequently the fine boundaries can themselves open.  
While the detailed structure depends on parameters in the model, generically we expect 
an incomplete devil's staircase to be realized.  In the figure,we have explicitly
drawn the opening of the[11]-[12] boundary but the other boundaries will behave
similarly.}
\label{fig:staircase}
\end{center}}
\end{figure}

We will also find that at higher orders, the individual boundaries of 
the [1n] sequence will themselves open revealing finer phase 
boundaries, which themselves can open.  This leads to the generic 
phase diagram sketched in Fig.~\ref{fig:staircase}.  
The steps in the [1n] sequence that are stabilized depend on the values of 
parameters in the Hamiltonian.  However, the conclusion of arbitrarily long
periods being realized is robust.  The fine structure of how the [1,n-1]-[1n] 
boundaries open is less certain because the dependence on parameters is more intricate 
and increasingly complicated resonances need to be accounted for.  
However, general arguments indicate that the boundaries will open 
and even periods incommensurate with the lattice\cite{footnote:incomm} 
are likely to occur in the model, though such states will not be seen at any 
finite order of our perturbation theory.

In the terminology of commensurate-incommensurate phase transitions, 
the [1n] sequence forms a (harmless) staircase with a ``devil's 
top-step''\cite{fishselke80,bak82}.  With the openings of these boundaries, beginning in the 
[11] state and moving left in Fig.~\ref{fig:staircase} for $t\neq 0$, 
the system traverses an incomplete devil's staircase of periodic 
states.  The subsequent steps in the staircase have progressively 
smaller tilts culminating in the flat columnar state.  The phase 
boundaries are first order transitions.  This phase diagram is similar 
to what is seen in the classical ANNNI model, where the transitions 
are driven by thermal fluctuations.

We make two remarks before launching into the calculation.  First, 
when we refer to the ``[11] phase'' (for example) what we 
precisely mean is that in this region, the ground state wavefunction is a 
superposition of dimer coverings that has relatively large overlap 
with the state in Fig.~\ref{fig:ideal}a and much smaller overlaps (of 
order $t^{2}$, $t^{4}$ etc.\ ) with excited states obtained by acting 
on Fig.~\ref{fig:ideal}a with Eq.~\ref{eq:V}.  The coefficients follow 
from perturbation theory.  Second, since Fig.~\ref{fig:staircase} is 
obtained using perturbation theory, we can be confident that this 
describes our system only in the limit where $t$ is small.  In the 
classical ANNNI model, numerical evidence indicates that as the small 
parameter (the temperature in that case) is increased, the phase 
boundaries close into Arnold tongue structures.  We do not currently 
know if this will occur in our model as $t$ increases.

%\vspace{1.4cm}
\section{Details}
\label{sec:details}
In this section, we construct a Hamiltonian using the strategy 
outlined in the previous sections.  The Hamiltonian $H=H_{0}+tV$ 
consists of a diagonal term $H_{0}$ and an off-diagonal term $tV$, 
which we treat perturbatively in the small parameter $t$.   

\subsection{Parent Hamiltonian}
Our parent Hamiltonian $H_{0}$ is the following operator:
\beq
\scalebox{0.43}{\includegraphics{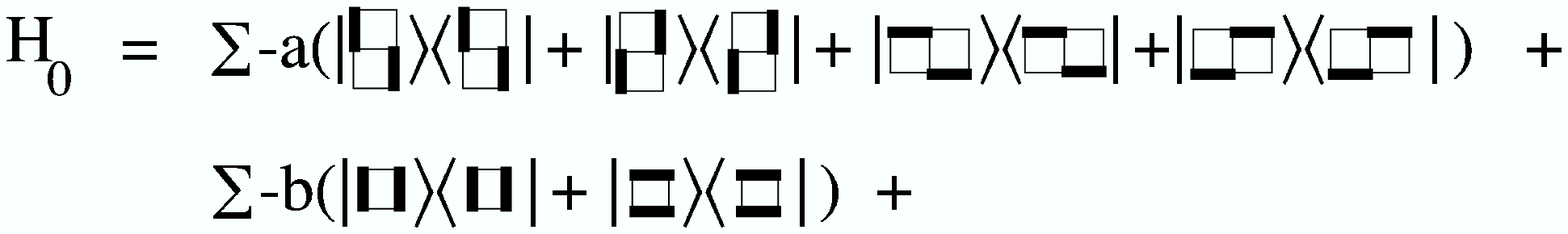}}\nonumber
\eeq
\vspace{-0.95cm}
\beq
\scalebox{0.43}{\includegraphics{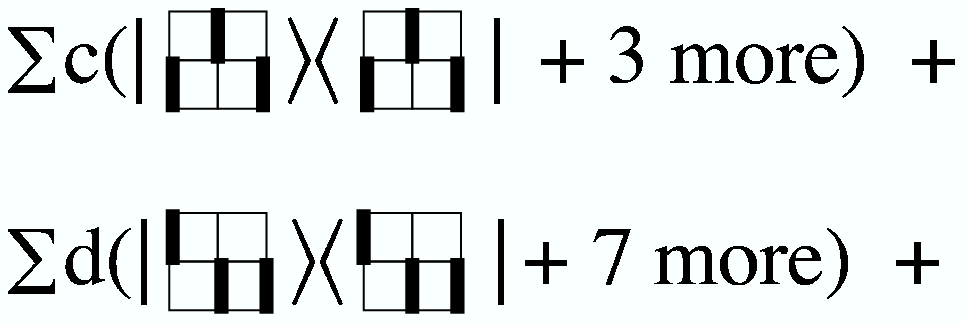}}\hspace{1.9cm}\nonumber
\eeq
\vspace{-0.95cm}
\beq
\scalebox{0.43}{\includegraphics{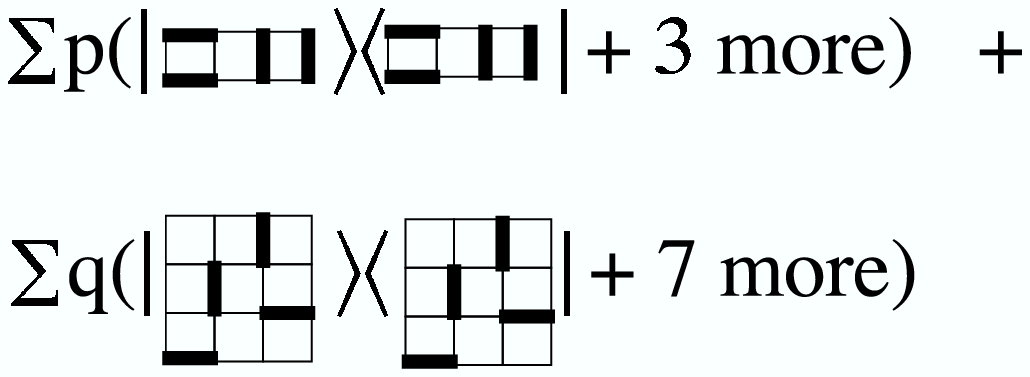}}\hspace{2cm}
\label{eq:H0}
\eeq
The coefficients $a$, $b$, $c$, $d$, $p$, and $q$ are 
positive numbers.  The symbols used in Eq.~\ref{eq:H0} are
projection operators referring to configurations of dimers on clusters 
of plaquettes and the sums are over all such clusters.  The notation 
``3 more'', etc.\ refers to the given term as well terms related 
to it by rotational and/or reflection symmetry; in terms $a$ and $b$ 
these other terms are explicitly written.  Notice that this 
Hamiltonian is a sum of local operators and does not break any 
symmetries of the underlying square lattice.

\begin{figure}[t]
{\begin{center}
\includegraphics[width=3in]{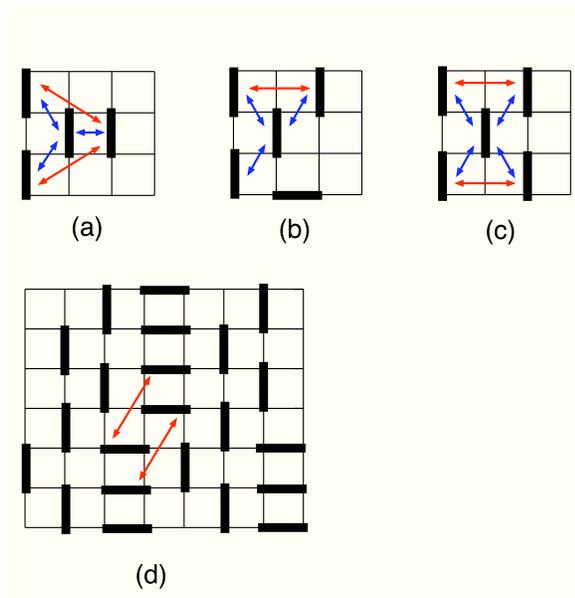}
\caption{(Color online) (a) and (b) are the two ways in which a dimer
can participate in three attractive bonds.  (c) is the 
one way in which a dimer can participate in four attractive bonds.  
The attractive bonds are shown by the blue (dark gray) arrows.  However, these 
configurations also involve repulsive interactions, which are shown 
in red (light gray), from terms $c$ and $d$ in the Hamiltonian.  (d) is an example 
of a state where every  
where every dimer has only two attractive 
bonds but with some dimers the two bonds are of different types.  
These ``kinks'' in the staggered domain walls involve an energy cost 
from term $d$ as indicated by the red (light gray) arrows.}
\label{fig:34}
\end{center}}
\end{figure}

Terms $a$ and $b$ are attractive interactions favoring staggered and columnar 
dimer arrangements respectively and we study Eq.~\ref{eq:H0} near  
$a=b$.  Terms $c$ and 
$d$ are repulsive interactions and if $c,d\geq a,b$, 
the dimers prefer domain wall patterns 
(Fig.~\ref{fig:generic}).  Terms $p$ and $q$ are repulsive interactions which 
determine the phases on the staircase.  If 
these terms are sufficiently large\cite{infinite} compared to $a,b,c$ and $d$, 
the staircase will include phases with arbitrarily long periods.     

We begin by showing that when $a=b$, the ground states of $H_{0}$ are 
the domain wall states of Fig.~\ref{fig:generic}.  We do this by showing that 
competitive states must have higher energy.  In the domain wall 
states, every dimer participates in exactly two attractive interactions and no 
repulsive interactions.  The only way to achieve a lower energy is for 
some dimers to participate in three or four attractive 
interactions.  This involves local dimer patterns of the form shown in 
Fig.~\ref{fig:34}abc.  In 
Fig.~\ref{fig:34}a, the central dimer participates in one columnar 
and two staggered interactions but also two repulsive interactions from 
term $d$ in Eq.~\ref{eq:H0}.  Similarly, 
Figs.~\ref{fig:34}b and \ref{fig:34}c show that if a dimer 
participates in more than two staggered interactions, the extra 
bonds are penalized by term $c$.  If we require $c,d\geq a,b$, these 
patterns will result in higher energy states as the
repulsive terms nullify the advantage of having extra attractive bonds.  
This also explains why in Fig.~\ref{fig:generic}, 
the staggered strips have unit width and the staggered and columnar dimers 
have opposite orientation.

Of the states where every dimer has two attractive 
bonds, the states where some dimers have two bonds of different type
will also have higher energy as shown in Fig.~\ref{fig:34}d.  Of the 
remaining states, it is readily seen that states where every dimer 
participates in either two $a$ bonds or two $b$ bonds, and where there 
are some $b$ bonds, must be of the domain wall form.  The only other 
possibility is the ``herringbone state'' where every dimer has two $a$ 
bonds (see Fig.~\ref{fig:phase0}).  The latter states are part of the 
degenerate manifold at $a=b$ but are dynamically inert because in this 
state it is not possible to locally manipulate the dimers (without 
violating the hard-core constraint). This establishes that when $a=b$, the ground states have the domain 
wall form.  It is also clear that when $a<b$, the system will maximize 
the number of $b$ bonds and when $a>b$, the number 
of $a$ bonds.  Therefore, we obtain the zero temperature phase 
diagram in Fig.~\ref{fig:phase0}. 

 It is useful to see this formally by calculating the energy of each domain wall state.
For concreteness, we assume the translational 
symmetry is broken in the $x$ direction.  A configuration with 
$N_{s}$ staggered strips and $N_{c}$ columnar strips has energy:
\bea
E(N_{s}, N_{c})&=& -a N_{y}N_{s} -b N_{y}N_{c}\nonumber\\&=&
-b\frac{N_{y}N_{x}}{2}+(b-a)N_{y}N_{s} 
\label{eq:energy0}
\eea
where $N_{x}$ and $N_{y}$ are the dimensions of the lattice (the 
lattice spacing is set to unity).  We have used the relation 
$N_{s}+N_{c}=\frac{N_{x}}{2}$.  When $a=b$, the 
domain wall states are degenerate and the energy scales with the total 
number of plaquettes $N_{x}N_{y}$.  If $a<b$, the system prefers the 
minimal number of staggered strips, which is the columnar state.  If 
$a>b$, the herringbone configuration has lower energy than any 
domain wall state.
Notice that all of these states are separated by a gap of order 
$a$ or $b$ from the nearest excited states obtainable by 
local manipulations of dimers.  Since we will be 
interested mainly in the difference $a-b$, the individual size of $a$ 
(or $b$), which sets the scale of this gap, can be made arbitrarily large.   

\subsection{Perturbation}
We now consider the effect of perturbing the parent Hamiltonian 
(\ref{eq:H0}) with the non-diagonal resonance term given in 
Eq.~\ref{eq:V}:
\begin{equation}
\scalebox{0.5}{\includegraphics{v.eps}}\\
\label{eq:V2}
\end{equation}
We assume $t<<1,a,b$ and consider $t$ as a
small parameter in perturbation theory.  We examine how the 
degeneracies in the $t=0$ phase diagram (Fig.~\ref{fig:phase0}) get 
lifted when $t\neq 0$.  The technical complications of degenerate 
perturbation theory do not arise because different domain wall 
states are not connected by a finite number of applications of 
this operator.  Eq.~\ref{eq:V2} is equivalent to two applications of the 
familiar two-dimer resonance of Rokhsar-Kivelson.  The mechanism
we now discuss can, in principle, be made to work for even 
this two-dimer term, but there are additional subtleties which will be 
mentioned.  

\subsubsection{Second order}
An even number of applications of operator \ref{eq:V2} are 
required to connect a domain wall state back to itself.  
Therefore, to linear order in $t$, the energies of these states are 
unchanged.  To second order in $t$, the energy shift of state 
$|n\rangle$ is given by:
\bea
E_{n}&=&
\epsilon_{n}-t^{2}\sum_{m}' \frac{V_{nm}V_{mn}}
{\epsilon_{m}-\epsilon_{n}}+O(t^{4})
\label{eq:pert2}
\eea
$\epsilon_{n}$ is the unperturbed energy of state $|n\rangle$ as 
given by Eq.~\ref{eq:energy0}.  The primed summation is over all dimer 
coverings except the original state $|n\rangle$.  The terms in the 
sum which give nonzero contribution correspond to states connected to the 
initial state by a single flipped cluster.  These terms may be 
interpreted as virtual processes taking the initial state to and from higher 
energy intermediate states, which may be viewed as quantum 
fluctuations of the staggered lines.  

\begin{figure}[ht]
{\begin{center}
\includegraphics[width=3in]{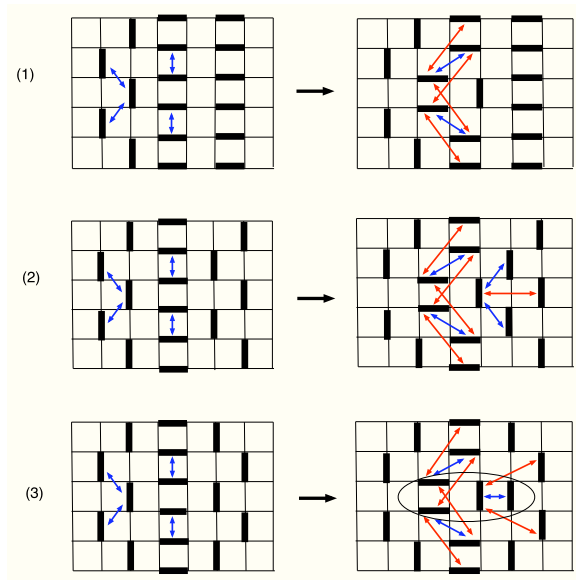}
\caption{(Color online) The three types of intermediate states obtained by 
acting once with the perturbation \ref{eq:V2} on the domain wall states. 
The blue/dark gray (red/light gray) arrows denote attractive (repulsive) interactions that are 
present in the initial (left) or excited (right) state but not both.  The 
circled cluster of the excited state of (3) is another such interaction.
(1) is a staggered line next to a columnar region of greater than unit width.  
(2) and (3) are staggered lines separated by one columnar strip 
from another staggered line of the same and opposite orientation 
respectively.  Notice that relative to the excited state of (1), the excited 
state of (2) has 2 additional attractive $a$ bonds and 2 additional 
repulsive $c$ bonds.  Similarly, excited state (3) has 1 additional 
attractive $b$ bond, 2 additional repulsive $d$ bonds, and a 
repulsive $p$ interaction.  Since $c,d\geq a,b$, cases (2) and (3) 
involve higher energy intermediate states than (1).}
\label{fig:singleflip}
\end{center}}
\end{figure}

The resonance energy of a particular staggered line depends on how the dimers 
next to the line are arranged.  Fig.~\ref{fig:singleflip} shows the three 
possibilities that give denominators that 
may enter Eq.~\ref{eq:pert2}: The staggered line may be of type (1) next to a 
columnar region of greater than unit width 
(Fig.~\ref{fig:singleflip}-1), or separated by one columnar strip 
from another staggered line of type (2), having the same orientation (Fig.~\ref{fig:singleflip}-2) 
or type (3), having opposite (Fig.~\ref{fig:singleflip}-3) orientation.  
From the figure, we see that cases (2) and (3) involve higher energy 
intermediate states than case (1) but allow for a denser packing of 
lines.  

If $p$ is sufficiently large, states with type (3) lines will 
be disfavored as ground states.  Ignoring such states, we update
Eq.~\ref{eq:energy0} to include second order corrections.  For  
convenience, we set $c=a$ which removes the distinction between 
cases (1) and (2), 
\bea
E(N_{s})&=&
-b\frac{N_{y}N_{x}}{2}+(b-a-\frac{t^{2}}{4d+2b})N_{y}N_{s}\nonumber\\
\label{eq:energy2}
\eea
We may use this to update the ground state phase diagram.
If $b>a+\frac{t^{2}}{4d+2b}$, the system is optimized when 
$N_{s}=0$, which is the columnar phase.  If $b<a+\frac{t^{2}}{4d+2b}$, 
the best domain wall state is the one with maximal staggering but 
without case (3) lines.  This corresponds to the [11] state 
(Fig.~\ref{fig:ideal}a) in which every staggered strip has the same orientation.
As $a-b$ is further increased, the herringbone state will eventually 
be favored.  The boundary between the [11] and herringbone state may 
be determined by comparing energies.  The energy of the [11] state is:
\bea
E_{[11]}=(-b-a-\frac{t^{2}}{4d+2b})\frac{N_{x}N_{y}}{4}+O(t^{4})
\eea
while the herringbone state has energy $E_{h}=-a\frac{N_{x}N_{y}}{2}$.
From this, it follows that the [11] state will be favored when:
\bea
a-\frac{t^{2}}{4d+2b}<b<a+\frac{t^{2}}{4d+2b}
\eea
while the herringbone state will occur when 
$b<a-\frac{t^{2}}{4d+2b}$.  

Therefore, up to corrections of order $t^{4}$, the system has the 
phase diagram shown in Fig.~\ref{fig:phase2}.  Because the coefficient of 
$N_{s}$ in Eq.~\ref{eq:energy2} is 
zero on the [11]-columnar boundary, we have that the [11] state, 
columnar state, and any domain wall state with intermediate tilt 
(that contains only lines of type (1) and (2)) are degenerate on the 
boundary.  In contrast, on the [11]-herringbone boundary, only the two states 
are degenerate.

\begin{figure}[t]
%\vspace{-2.5cm}
{\begin{center}
\includegraphics[width=0.37\textwidth]{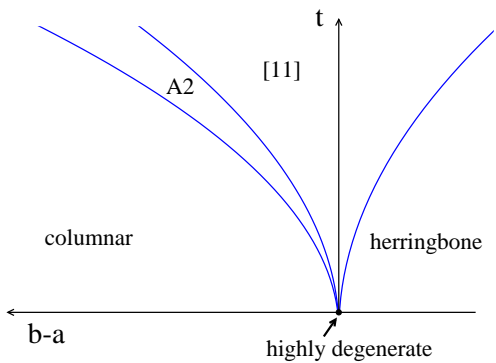}
%%%\vspace{-1cm}
%\includegraphics[width=3in]{stairs2B.eps}
\caption{(Color online) Second order phase diagram when $c>a$.  The [11] phase has 
width $\sim t^{2}$ and the $A_{2}$ phase has 
width $\sim (c-a)t^{2}$.  On the [11]-$A_{2}$ and the 
$A_{2}$-columnar boundaries, intermediate domain wall states are 
degenerate as described in the text.}
\label{fig:phase2b}
\end{center}}
\end{figure}

In Appendix \ref{app:ca}, the more general case, where $c>a$, is 
discussed and the resulting phase diagram is shown in 
Fig.~\ref{fig:phase2b}.  An additional phase is 
stabilized in a region of width $\sim (c-a)t^{2}$ (assuming $|c-a|$ is finite) 
between the [11] and columnar states.  In this new phase, labelled $A_{2}$, any state where 
adjacent staggered lines are separated by two columns, including the 
[12] state (Fig.~\ref{fig:ideal}b), is a ground state.  These are the 
states which maximize the number of type (1) staggered lines 
(Fig.~\ref{fig:singleflip}-1).  On 
the [11]-$A_{2}$ boundary, intermediate states where adjacent 
staggered lines are separated by one or two columns (and where there 
are no type (3) lines) are degenerate.    
On the $A_{2}$-columnar boundary, states where 
adjacent staggered lines are separated by at least two columns are 
degenerate.       

Fig.~\ref{fig:phase2b} may be understood intuitively by noting that 
at $a=b$ and $t=0$, a staggered strip has the same energy as a columnar strip.
The resonance terms lower the effective energy of a staggered strip 
and since the [11] state involves the most staggered strips, its 
energy will be lowered the most.  As $b$ increases to the point where 
the degeneracy between columnar strips and type (2) lines is restored, 
the system will prefer to maximize the number of type (1) lines, which 
are individually more stable but loosely packed.  This is the 
transition to the $A_{2}$ phase.  As $b$ is increased further, the 
degeneracy between columnar strips and type (1) lines is restored and 
there is a transition to the columnar state.  Note that if $|c-a|$ is 
very large, [11]-$A_{2}$ boundary can occur in the $a>b$ region.   

Before proceeding with the calculation, we clarify two points of 
potential confusion.  First, when we say the ``[11] 
state is stabilized'' over part of the phase diagram, what we precisely
mean is that the ground state wavefunction is a superposition of dimer 
coverings that has relatively large overlap with
the literal [11] state of Fig.~\ref{fig:ideal}a and much smaller 
overlaps (of order $t^{2}$) with the excited states obtained by acting 
on the [11] state once with the perturbation \ref{eq:V2} (Fig.~\ref{fig:flip}).  
We use the notation [11] to denote both the perturbed wavefunction, which is an 
eigenstate of the perturbed Hamiltonian and the literal [11] state, which is an 
eigenstate of the unperturbed Hamiltonian.  Second, the phase 
boundaries are based on a competition 
between a resonance term, which is a quantum version of ``entropy'', and 
{\em part} of the zeroeth order piece which, continuing the classical 
analogy, is like an internal energy.  This does not contradict the spirit of 
perturbation theory because the {\em full} zeroeth order term, 
$-b\frac{N_{y}N_{x}}{2}+(b-a)N_{y}N_{s}$, is always larger than the 
second order correction. 

A similar phase diagram will occur (though at fourth order in perturbation theory) if we use the 
two-dimer move of Rokhsar-Kivelson.  However, the bookkeeping will be more complicated
because resonances will be able to originate in the interior of the columnar regions instead of 
just at that the columnar-staggered boundaries.  This may be compensated by tuning 
$b$, which will merely move the boundaries, or by adding appropriate (local) repulsive terms to 
the parent Hamiltonian.      

\subsubsection{Fourth order}
\label{sec:fourth}

We concentrate on Fig.~\ref{fig:phase2b} as it is
more generic than the fine-tuned $c=a$ case of 
Fig.~\ref{fig:phase2}.  In either case, we expect the degeneracies on 
the phase boundaries to be partially lifted by considering higher orders in 
perturbation theory.  In this section, we focus on the 
$A_{2}$-columnar boundary, where adjacent lines are 
separated by at least two columnar strips.  

We return to 
the perturbation series for the energy (Eq.~\ref{eq:pert2}), 
this time keeping terms up to fourth order in the small parameter.  
\bea
E_{n}&=&
\epsilon_{n}-t^{2}\sum_{m}'\frac{V_{nm}V_{mn}}{\epsilon_{m}-\epsilon_{n}}
\nonumber\\
&-& t^{4}\Bigl[\sum_{ml}'\frac{V_{nm}V_{ml}V_{lk}V_{kn}}
{(\epsilon_{m}-\epsilon_{n})(\epsilon_{l}-\epsilon_{n})(\epsilon_{k}-\epsilon_{n})}
\nonumber\\&-&\sum_{ml}'\frac{V_{nm}V_{mn}V_{nl}V_{ln}}{(\epsilon_{m}-\epsilon_{n})^{2}
(\epsilon_{l}-\epsilon_{n})
}\Bigr]+ O(t^{6})
\label{eq:pert4}
\eea
As usual, the primes denote that the sum is over all states except 
$|n\rangle$.  The two fourth order terms correspond, in 
conventional Rayleigh-Schrodinger perturbation theory, to the 
corrections to the energy expectation value $\langle\psi|H|\psi\rangle$
and wavefunction normalization $\langle\psi|\psi\rangle$.  We will 
refer to the corresponding terms as ``energy'' and ``wavefunction'' 
terms.  As in the second order case, we may view the terms in 
the fourth order sums as virtual resonances connecting the initial state to 
itself via a series of high energy intermediate states.  For this 
reason, we will refer to the summands as (fourth-order) ``resonances''.

\begin{figure}[ht]
{\begin{center}
\includegraphics[width=3in]{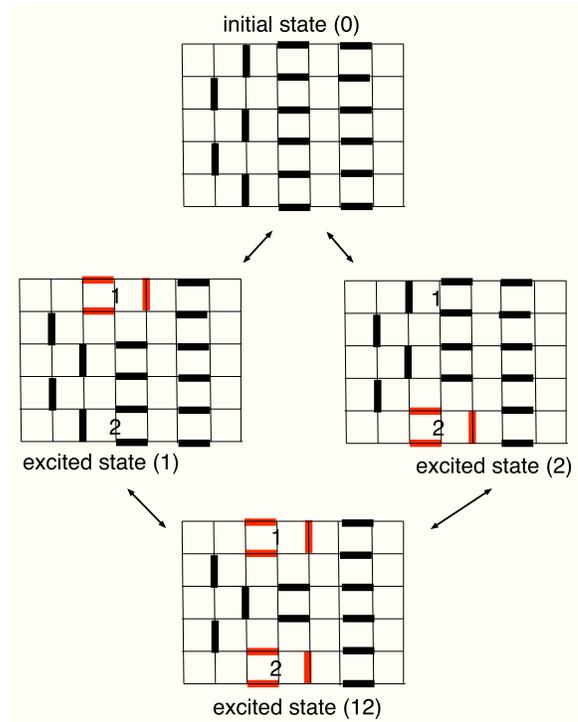}
\caption{(Color online) Most of the fourth order terms in Eq.~\ref{eq:pert4} 
involve ``disconnected'' clusters of dimers.  In this figure, 
the perturbation connects the initial state (0) to excited states, labelled 
(1) and (2), depending on whether cluster 1 or 2 has been flipped.  Acting 
again with the perturbation connects to an excited 
state, labeled (12), where both of these clusters are flipped.  
Acting two more times with the perturbation brings us back to the 
initial state (0) via either of excited states (1) or (2).  Such 
terms are called disconnected because there are no interactions (in 
Eq.~\ref{eq:H0}) between the dimers of clusters 1 and 2.  The figure depicts a 
particular resonance from the energy term in Eq.~\ref{eq:pert4}.  
There is an analogous contribution from the wavefunction term which is 
a product of the second order processes $(0)\rightarrow(1)\rightarrow(0)$
and $(0)\rightarrow(2)\rightarrow(0)$.  While the number of such disconnected 
terms scales as $N^{2}$, where $N=L_{x}L_{y}$ is the system size, 
these resonances do not contribute to the energy because the 
contributions from the energy and wavefunction terms exactly cancel.}
\label{fig:disconnect}
\end{center}}
\end{figure}

Most terms in the double sum in Eq.~\ref{eq:pert4} correspond to 
resonances between disconnected clusters 
(Fig.~\ref{fig:disconnect}).  Referring to the figure, we use the term 
``disconnected'' to indicate that there are no interaction terms connecting 
the dimers of clusters 1 and 2.  While the number of such resonances 
scales as the square of the system size, the contributions from the 
energy and wavefunction terms in Eq.~\ref{eq:pert4} precisely cancel 
for these disconnected clusters.  The details of this cancellation 
are discussed in Appendix \ref{app:disconnect}.     

\begin{figure}[ht]
{\begin{center}
\includegraphics[width=3in]{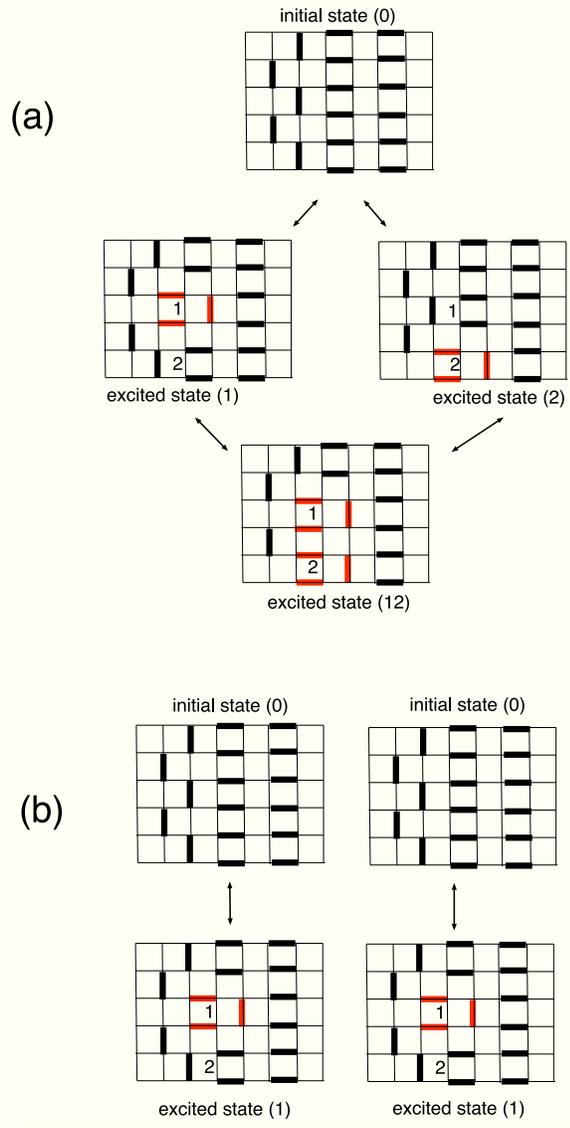}
\caption{(Color online) These are examples of fourth order self-energy resonances.  
Each resonance is confined to a single line and the number of 
resonances in the system is proportional to the number of lines.
In resonances such as (a), there are interactions 
connecting dimers of two flipped clusters on the same line.
Terms such as (b) arise 
from the wavefunction term in Eq.~\ref{eq:pert4} but have no analogous 
processes in the energy term to cancel against.}
\label{fig:four_line}
\end{center}}
\end{figure}

The remaining fourth order resonances are extensive in number and may be 
grouped into three categories.  In the first category, shown in 
Fig.~\ref{fig:four_line}, are resonances associated with a single 
staggered line and the number of such resonances in the system is proportional 
to the number of lines.  We refer to these resonances as 
``self-energy corrections''.

\begin{figure}[ht]
{\begin{center}
\includegraphics[width=3in]{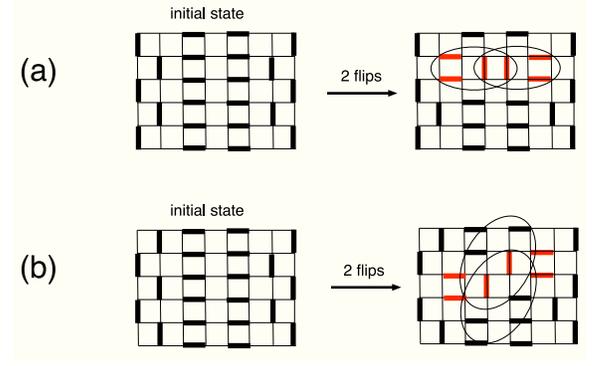}
\caption{(Color online) These are examples of fourth order resonances which are 
effective interactions between lines.  At fourth order, such interactions are 
possible only when lines are separated by two or fewer columns.  On 
the $A_{2}$-columnar boundary, resonances (a) and (b) are the only 
processes to consider.  These involve terms $p$ and $q$ in the 
Hamiltonian, as indicated by the circles.}
\label{fig:four_suppress}
\end{center}}
\end{figure}

In the second category, shown in Fig.~\ref{fig:four_suppress}, there are 
resonances that contribute to the effective interactions between adjacent lines.
These resonances occur only in states where lines are
separated by two or fewer columnar strips.  Because we are interested 
in the $A_{2}$-columnar boundary, the only processes to 
consider are the ones shown in the figure.  The purpose of terms $p$ 
and $q$ in Eq.~\ref{eq:H0} is to control the processes in 
Figs.~\ref{fig:four_suppress}a and b respectively.  The net 
contribution of these resonances involve both energy and wavefunction 
terms.  For example, the contribution of Fig.~\ref{fig:four_suppress}b 
to the energy is:
\beq
e=\frac{2t^{4}}{2(4d+2b)^{3}}\Bigl[1-\frac{1}{1+\frac{2q-a}{2(4d+2b)}}\Bigr]
\label{eq:e}
\eeq
and likewise for Fig.~\ref{fig:four_suppress}a (replace $2q-a$ with 
$2p-b$).  If $q>a/2$, the net contribution of this resonance is repulsive.

\begin{figure}[ht]
{\begin{center}
\includegraphics[width=3in]{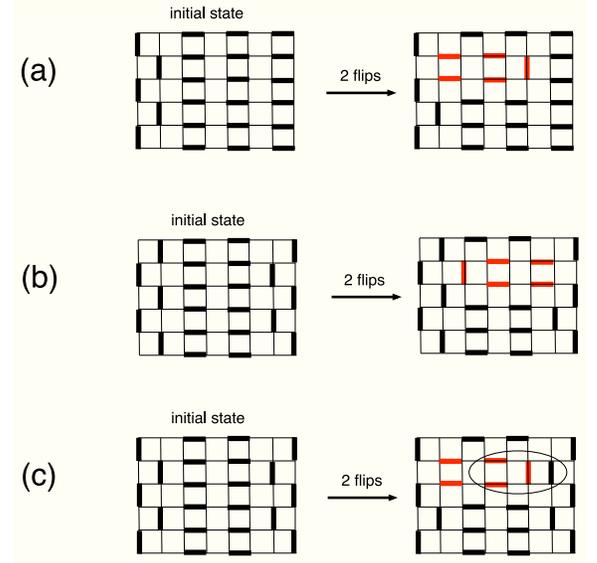}
\caption{(Color online) These are the long fourth order resonances which occur in 
states where lines are separated by two or more columnar strips.  
These processes are always stabilizing though the amount depends on 
the environment of the line as in the second order (see 
Fig.~\ref{fig:singleflip}).  The resonance in (a) is strongest because 
it involves the lowest energy intermediate state but (b) and (c) allow 
for a denser packing of lines.  Resonance (c) is especially 
suppressed due to term $p$ in Eq.~\ref{eq:H0}.}
\label{fig:four_long}
\end{center}}
\end{figure}

The third type of resonance is the long resonance, shown 
in Fig.~\ref{fig:four_long}.  These resonances occur in the 
energy term of Eq.~\ref{eq:pert4} but do not have corresponding pieces
in the wavefunction term.  Therefore, these resonances will always 
lower the energy though, as the figure indicates, the precise amount 
depends on the way the lines are spaced.

An immediate implication of these resonances is the lifting of the 
degeneracy of the $A_{2}$ phase.  All of these states have the same 
number of lines so will receive the same self-energy contribution 
(Fig.~\ref{fig:four_line}).  If we choose $p,q$ large compared to 
$a,b,c$ and $d$, then the repulsive contribution from the interaction 
resonances (Fig.~\ref{fig:four_suppress}) is essentially determined 
by the wavefunction term, which is the same for all of the $A_{2}$ 
states.  The degeneracy is broken by the long resonances because in 
the [12] state, only Fig.~\ref{fig:four_long}b processes occur 
while in the other $A_{2}$ states, some of the resonances are the 
suppressed Fig.~\ref{fig:four_long}c variety.  Therefore, what was 
seen as merely an $A_{2}$ phase at second order is revealed, on 
closer inspection, as a [12] phase.

To investigate the degeneracy of what we now recognize as the [12]-columnar 
boundary, it is useful to update Eq.~\ref{eq:energy2} to include 
fourth order corrections:
\bea
E &=& 
-b\frac{N_{y}N_{x}}{2}+(b-a-\frac{t^{2}}{4d+2b}+\alpha 
t^{4})N_{y}N_{s}\nonumber\\
&-&\beta t^{4}N_{y}N_{sa}-\gamma t^{4} N_{sb}+O(t^{6})
\label{eq:energy4}
\eea
where $N_{s}$ is the total number of staggered lines and $N_{sa(b)}$ is 
the number of staggered lines having the environment of 
Fig.~\ref{fig:four_long}a(b).  We ignore states with arrangements like 
Fig.~\ref{fig:four_long}c since they are disfavored as ground states.
 $\alpha$ is a 
constant, which may be calculated but whose value is unimportant, 
containing the contribution of fourth order self energy terms.  
$\beta=\frac{1}{(4d+2b)^{2}(2(4d+2b)+4(d-a))}>0$ is the 
contribution of the most favorable long fourth order resonances 
(Fig.~\ref{fig:four_long}a) 
whose number is proportional to $N_{sa}$.  
$\gamma=\frac{1}{(4d+2b)^{2}(2(4d+2b)+4(d-a)+2(c-a))}-2e$ ($e$ given 
by Eq.~\ref{eq:e}) is proportional to $N_{sb}$ and includes the 
contributions of Figs.~\ref{fig:four_suppress}b and \ref{fig:four_long}b.  
Note that $\gamma<\beta$ and the sign of $\gamma$ is determined by 
the size of $q$.  For convenience, we assume $q$ large enough that 
$\gamma<0$. 

We may use Eq.~\ref{eq:energy4} to correct the phase diagram.
Similar to the second order case, as $b$ is increased, the extra stability of 
the staggered strips in the [12] state becomes eventually balanced by the 
zeroeth order energy of the columnar strips.  When this occurs, the 
system will prefer a state with fewer lines that are individually more 
stable.  In particular, the states we may label $A_{3}$, where 
adjacent lines are separated by three columns, maximize the number of 
favorable long resonances (Fig.~\ref{fig:four_long}a) without incurring 
any of the repulsive fourth order penalties (i.\ e.\ the analog of 
Fig.~\ref{fig:four_suppress} would be a disconnected resonance).  
In the next section, higher order perturbation theory will show that this $A_{3}$ phase is 
actually a [13] phase (Fig.~\ref{fig:ideal}c) so we 
begin using the [13] label immediately.  As $b$ is increased further, 
the system will enter the columnar state.  

\begin{figure}[t]
%\vspace{-2.5cm}
{\begin{center}
\includegraphics[width=0.37\textwidth]{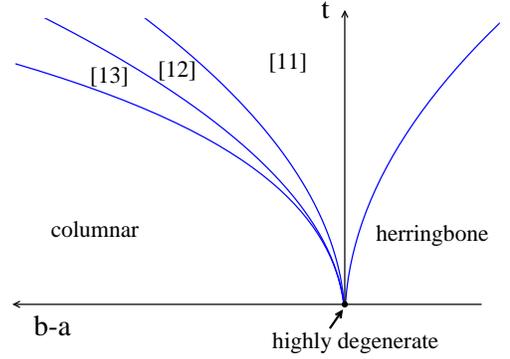}
%\includegraphics[width=3in]{stairs4B.eps}
%%%\vspace{-1cm}
\caption{(Color online) Fourth order phase diagram.  The new [13] state has width $\sim 
t^{4}$.}
\label{fig:phase4b}
\end{center}}
\end{figure}

The result is the phase diagram in Fig.~\ref{fig:phase4b}.
The phase boundaries are determined by comparing energies.
Ignoring corrections of order $t^{6}$, we have the following:
\bea
E_{[12]}=-b\frac{N_{y}N_{x}}{2}+(b-a-\frac{t^{2}}{4d+2b}+\alpha 
t^{4}-\gamma t^{4})\frac{N_{y}N_{x}}{6}\nonumber\\
\label{en12}
\eea
\bea
E_{[13]}=-b\frac{N_{y}N_{x}}{2}+(b-a-\frac{t^{2}}{4d+2b}+\alpha 
t^{4} - \beta t^{4})\frac{N_{y}N_{x}}{8}\nonumber\\
\label{en13}
\eea
\beq
E_{col}=-b\frac{N_{y}N_{x}}{2}
\eeq

Comparing these expressions, we obtain the updated phase diagram shown in 
Fig.~\ref{fig:phase4b}.  The [12] state is favored when:
\beq
b<a+\frac{t^{2}}{4d+2b}-\alpha t^{4}-(4|\gamma|+3\beta)t^4
\label{eq:boundary4}
\eeq

The [13] is favored when:
\bea
a&+&\frac{t^{2}}{4d+2b}-\alpha t^{4}-(4|\gamma|+3\beta)t^{4}< b\nonumber\\&<&
a+\frac{t^{2}}{4d+2b}-\alpha t^{4}+\beta t^{4}\nonumber\\
\eea

The columnar state is favored when:
\beq
b>a+\frac{t^{2}}{4d+2b}-\alpha t^{4}+\beta t^{4}
\eeq

On the [12]-[13] boundary, there is a degeneracy between intermediate 
states where adjacent lines are separated by either two or three 
columnar strips.  On the [13]-columnar boundary, there is a degeneracy 
between states where adjacent staggered lines are separated by 
at least two columns.  

\subsubsection{Higher orders and fine structure}
\label{sec:sixth}

The picture of Fig.~\ref{fig:phase4b} will be further refined 
by considering sixth order resonances and new phases 
will appear near the phase boundaries, in regions of width $~\sim t^{6}$, 
which is why they were missed at fourth order.  The most immediate 
consequence will be the lifting of the degeneracy of the $A_{3}$ states in 
favor of the state [13]. The latter state, in comparison with the other $A_{3}$ 
states, is both stabilized maximally by the sixth order analog of 
Fig.~\ref{fig:four_long} and, if we choose $p>q$, destabilized minimally
by the sixth order analog of Fig.~\ref{fig:four_suppress}.  
The [13]-columnar phase boundary will open to reveal the [14] phase 
(Fig.~\ref{fig:ideal}d)\cite{footnote:14}, in which the number of 
favorable long resonances, the sixth order analogs of 
Fig.~\ref{fig:four_long}a, is maximized and there are no repulsive 
contributions ( i.\ e.\ the sixth order analogs of 
Fig.~\ref{fig:four_suppress} will be disconnected terms if the lines 
are more than three columnar strips apart).    

\begin{figure}[ht]
{\begin{center}
\includegraphics[width=2.5in]{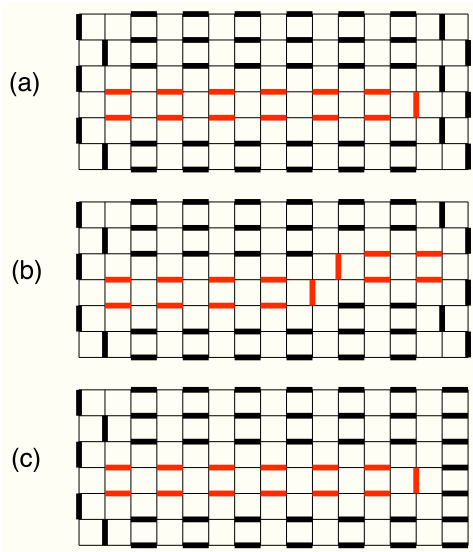}
\caption{(Color online) These are the 12-th order resonances which drive the transition between 
the [16] and [17] states.  Process (a) selects the [16] state from 
the others in the $A_{6}$ manifold, which were stabilized equally at 
10th order.  Process (b) destabilizes the [16] state near its 
columnar boundary in favor of the [17] state, which maximizes the number of 
most favorable resonances (c).}
\label{fig:nthorder}
\end{center}}
\end{figure}

The argument may be applied iteratively at higher orders in 
perturbation theory.  At 2n-th order, we may ask whether the 
[1n]-columnar boundary will open to reveal a new phase.  
The transition to less tilted states will be again be 
driven by processes that connect adjacent lines 
(Fig.~\ref{fig:nthorder}).  In the competitive states, 
adjacent lines are separated by at least n columnar strips so 
2n-th order resonances connecting the lines must be ``straight''.
This means that the complicated 
high order processes, including ``snake-like'' fluctuations that break the 
staggered lines, will simply change the self-energy of a staggered 
line and do not have any effect on the transition.  The [1n] phase will be 
destabilized by the process in Fig.~\ref{fig:nthorder}b which will 
overwhelm the stabilizing effect of Fig.~\ref{fig:nthorder}a due to 
combinatorics.  However, these repulsive processes will not contribute when 
the lines are separated by more than n columnar strips.  Therefore, 
the [1,n+1] phase, which maximizes the number of the long 2n-th 
order resonances (Fig.~\ref{fig:nthorder}c), will be stabilized in a region of 
width $~\sim t^{2n}$ between the [1n] and columnar phases.  
Therefore, we obtain the phase diagram of Fig.~\ref{fig:phaseN}.

This shows that states with arbitrarily long periods are stabilized 
without long range interactions or fine tuning (other than the requirements of perturbation
theory).  The situation will be similar if we were to use the two-dimer Rokhsar-Kivelson
resonance.  In this case, there will be contributions from resonances occurring only in the
columnar regions, which were ``inert" in our calculation.  These processes will amount to
self-energy corrections that just renormalize columnar energy scale $b$.  Also, additional local terms  (i.e. other than $p$ and $q$) may be required to realize the very high-order states because adjacent lines will be able to interact via intermediate states other than the ones shown in Fig.~\ref{fig:four_suppress}.  While we have not worked out the exact details of this case, we note that there are 
a finite number of such intermediates states so only a finite number of {\em local} terms will be 
required.  In particular, we will {\em not} have to add longer terms at each order in perturbation
theory.  

So far, we have concentrated on the boundary with the columnar phase but we may also 
ask whether a similar lifting may occur on the other phase 
boundaries.  We consider the [11]-[12] boundary.  Both of these 
phases are stabilized at second order and occupy regions of width 
$~\sim t^{2}$ in the phase diagram (assuming $|c-a|\gg t^{2}$).  On 
their boundary, all states where staggered lines are separated by 
either one or two columnar strips are degenerate to second order.  We 
investigate the effect of fourth order resonances on this boundary.  

\begin{figure}[ht]
{\begin{center}
\includegraphics[width=3in]{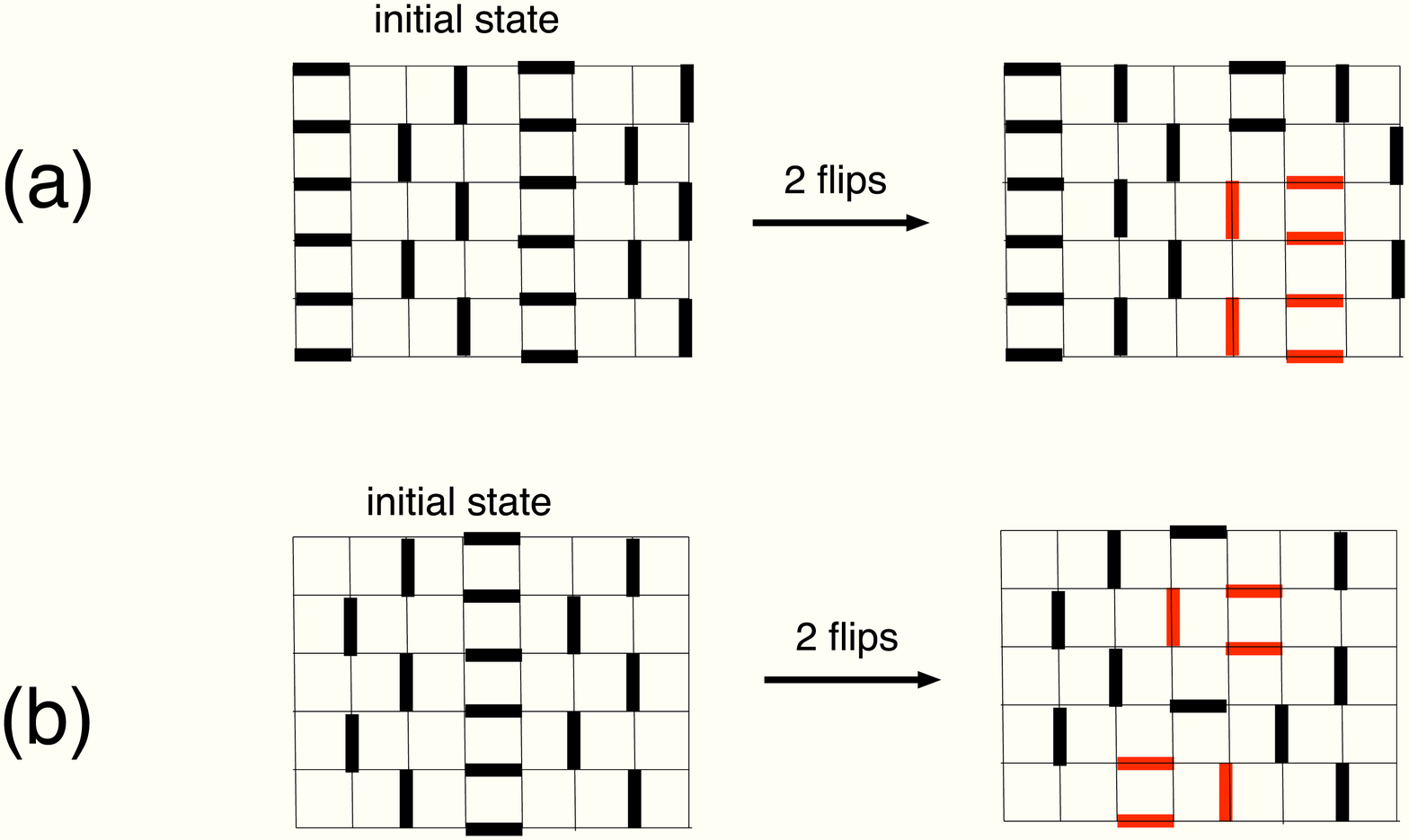}
\caption{(Color online) These are fourth order processes available between two 
staggered lines separated by a single columnar strip.}
\label{fig:boundary1112}
\end{center}}
\end{figure}

\begin{figure}[ht]
{\begin{center}
\includegraphics[width=2in]{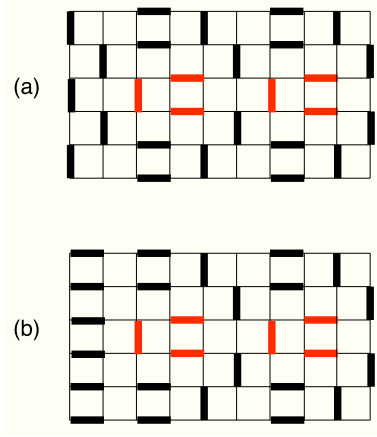}
\caption{(Color online) These are fourth order processes which can occur if a [11] 
region is next to (a) another [11] region or (b) a [12] region.  Both 
of these resonances contribute to the energy with a negative sign because the intermediate state 
involves two fewer repulsive $c$ bonds than if the flipped clusters 
were farther apart.}
\label{fig:boundary1112b}
\end{center}}
\end{figure}

We need to consider not only the resonances presented in section 
\ref{sec:fourth} but also new fourth order processes which become 
available once we consider staggered lines that are one column apart.
These are shown in Figs.~\ref{fig:boundary1112} and 
\ref{fig:boundary1112b}.  The resonances in section \ref{sec:fourth} 
will stabilize (or destabilize -- the sign is not important) 
each boundary state by an amount proportional 
to the number of its ``[12] regions'', i.\ e.\ columnar regions that are 
two columns wide and have staggered lines on their boundary, while the resonances in Fig.~\ref{fig:boundary1112} 
will contribute an amount proportional to the number of ``[11] 
regions'', i.\ e.\ columnar regions that are one column wide.  If 
these were the only available processes, the [11]-[12] boundary would 
be shifted by $\sim t^{4}$ but the degeneracy on the boundary would 
remain.

The possibility of a new phase is determined by the resonances in 
Fig.~\ref{fig:boundary1112b}.  Both of these resonances have an overall stabilizing effect (the contributions to the energy are negative)
and depend on whether a [11] region is adjacent to another [11] 
region (Fig.~\ref{fig:boundary1112b}a) or a [12] region 
(Fig.~\ref{fig:boundary1112b}b).  Because $c>a$, resonance (b) is 
stronger than (a), since its intermediate state has lower energy, but requires 
a lower density of staggered lines.  If the net contribution of 
resonance (a) wins, then the degeneracy would be lifted in favor of 
the [11] state and the [11]-[12] boundary would be shifted again, but 
the degeneracy will only be between the two states.  
However, if $c$ is made sufficiently large\cite{foot_csize}, the 
contribution of resonance (a) tends to zero while (b) approaches a constant 
because the intermediate state can occur without involving $c$ bonds 
(i.\ e.\ the fourth order process where the left cluster is flipped 
first and last).  Therefore, there will be a new phase where resonance 
(b) is maximized.  This state is the [11-12] state where the label 
refers to the repeating unit
``one staggered strip followed by one columnar strip followed by
one staggered strip followed by two columnar strips''.

\begin{figure}[t]
{\begin{center}
\includegraphics[width=2in]{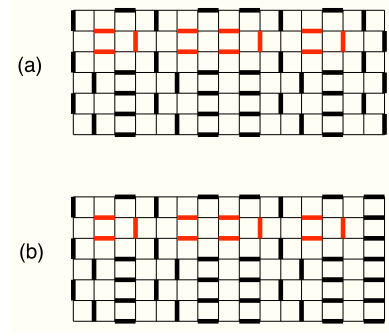}
\caption{(Color online) These are eighth order processes which can lead to a new 
phase between the [11-12] and [12] phases.  Resonance (a) stabilizes 
the [11-12] phase while (b) stabilizes the [11-(12)$^{2}$] phase.  In 
the limit where $c$ is large, resonance (b) is preferable.  The 
easiest way to see this is by setting $c=\infty$.  Then the energy 
term of resonance (a) and all of the wavefunction terms in resonances 
(a) and (b) will give zero because the 
intermediate states can not form without creating $c$ bonds.  
However, the intermediate state in 
resonance (b) does not involve $c$ bonds and can be obtained without 
creating $c$ bonds (i.\ e.\ in an 
eighth 
order process where the first and last two 
actions involve flipping the cluster on the right and the first 
cluster in the 
middle).  Therefore, the energy term of resonance (b) will give a 
stabilizing contribution.}
\label{fig:boundary1112c}
\end{center}}
\end{figure}

Continuing the line of thought, we may ask whether the [11-12]-[12] 
boundary will open at higher orders.  Sixth order resonances will 
shift the boundary but there are no processes which break the 
degeneracy.  However, at eighth order, there is a resonance which will 
favor a [11-(12)$^{2}$] phase (Fig.~\ref{fig:boundary1112c}).  

We can, in principle, investigate whether the [11-(12)$^{n}$] phases 
continue to appear when n is large and also whether the new  
boundaries themselves open to reveal even finer details.  
The same arguments will hold for all of the other boundaries in the [1n] 
sequence.  While the structure of our Hamiltonian allowed us to be 
definite regarding the [1n] sequence, it is more difficult to draw 
conclusions about the fine structure at high orders in perturbation 
theory because increasingly complicated resonances need to be 
accounted for.  Most of these resonances will stabilize the 
unit cells of the states on either side of the boundary so the net 
effect will be to move the boundary.  The more important terms, with 
respect to whether boundaries will open, are resonances associated 
with interfaces between regions with one or another unit cell.  
Even these terms can become complicated at very 
high orders in perturbation theory.  However, our arguments suggest 
that arbitrarily complicated phases can, in principle, be stabilized by 
going to an appropriate range in parameter space and/or adding 
additional local interactions.  Therefore, the most generic situation 
is an incomplete devil's staircase, as sketched in 
Fig.~\ref{fig:staircase}.  

\subsection{Order of the transitions between the $[1n]$ phases}

The boundaries between different modulated states are generically first-order. 
Intuitively, this is not surprising because of the topological property that ``protects" the states, namely that even with an infinite number of {\em local} dimer moves, it is \emph{impossible} to go from one state to the other since the states are in different topological sectors.
Therefore, we would not expect a transition to be driven by a growing correlation length.

Formally, the way we use to determine the order of the transitions that 
emerge in the system, is by calculating the first derivative of 
the ground-state energy on either side of the phase boundaries that we found.\cite{sachdev_book}
For example, we will treat explicitly the case of the transition between 
$[12]\rightarrow[13]$, even though the same line of argument applies to the other boundaries 
as well. The energies of the two states near the phase 
boundary, are given by Eq.~\ref{en12}, Eq.~\ref{en13} and the phase boundary is 
given by the following condition(as can be seen in 
Eq.~\ref{eq:boundary4}):
\beq
 b=a+\frac{t^{2}}{4d+2b}-\alpha t^{4} - (4|\gamma|+3\beta) t^4
\label{b4explicit}
\eeq
Let's consider the case where we approach the boundary from the $[12]$ side 
varying the variable $t$ but keeping $a,b$ constant. In the phase diagram 
Fig.~\ref{fig:phaseN}, we 'move' vertically down. The reason for choosing this direction is just clarity.  Let's call the point of the phase boundary where our path crosses, $A$, and the critical value of t, $t_c$ (coming from the solution of Eq.~\ref{b4explicit} for fixed values of $a,b,d$).

The energies of the two states at the phase boundary are exactly equal. Their derivatives are:

\begin{eqnarray}
\frac{\partial E_{[12]}}{\partial t}\Bigg|_{A^{+},t_c}&=&-(\frac{2t_c}{4d+2b}-4\alpha t_{c}^3 -4|\gamma| 
t_{c}^{3})\frac{N_{y}N_{x}}{6}\nonumber\\ &&
+O(t_{c}^{5})\label{ender1}\\
\frac{\partial E_{[13]}}{\partial t}\Bigg|_{A^{-},t_c}&=&-(\frac{2t_c}{4d+2b}-4\alpha t_{c}^{3}  + 4\beta t_{c}^{3})\frac{N_{y}N_{x}}{8}\nonumber\\ &&
+O(t_{c}^{5})
\label{ender2}
\end{eqnarray}
By Eqs.~\ref{b4explicit}, \ref{ender1} and \ref{ender2}, we have:
\begin{eqnarray}
\frac{\partial E_{[12]}}{\partial t}\Bigg|_{A^{+},t_c}-\frac{\partial E_{[13]}}{\partial t}\Bigg|_{A^{-},t_c} = \nonumber\\
 \underbrace{\left(\frac{t_c}{2(4d+2b)}-\frac{b-a}{t_c}\right)}_{<0 {\rm\; to\; order}\; O(t^2)} \frac{N_yN_x}{6} + O(t_{c}^5)
\end{eqnarray}
The derivatives are not equal along the phase boundary so the transition is discontinuous (first-order). It is clear that \emph{all} the phase transitions we found will also be discontinuous because the above discontinuity comes exactly from the contributions leading to the phase boundary's presence.

\section{Connections with frustrated Ising models}
\label{sec:spin}

A natural question to ask is whether the staircase presented above has any connection to the staircase of the 3D ANNNI model or the quantum analogs discussed in Ref.~\onlinecite{yeomans95}.  One of the main differences of the present work is the non-perturbative inclusion of frustration by considering hard-core dimers as the fundamental degrees of freedom.  In this sense, the present staircase differs from previous work similarly to how the fully frustrated Ising model differs from the conventional Ising model.  It is instructive to consider the mapping between dimer coverings and configurations of the fully frustrated Ising model on the square lattice (FFSI) in more detail.       

\begin{figure}[h]
\includegraphics[width=0.45\textwidth]{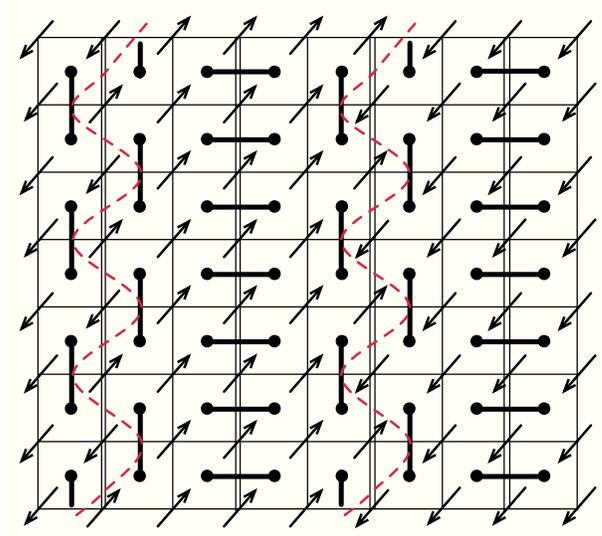}
\caption{(Color online) One of the four possible $[11]$ configuration in terms of Ising spins on the fully frustrated square lattice.  The hard-core dimer constraint corresponds to the requirement that the FFSI ground state has one ``unsatisfied" bond per plaquette.  The columnar-dimer regions correspond to ferromagnetic-Ising domains. The staggered-dimer strips correspond to the Ising domain walls, depicted by the red-colored (dashed) curved lines, which separate ferromagnetic domains of different orientation. Clearly, in the other equivalent configurations, even though they share the same principle of domain-wall competition, the separated regions are not ferromagnetic but one of several metamagnetic choices \cite{liebmann_book}.}
\label{fig:ffsi}
\end{figure}
The FFSI model can be described in terms of Ising degrees of freedom living on the square lattice. The main difference with the usual ferromagnetic Ising model is the following: Even though in the x-direction, all the bonds are ferromagnetic, in the y-direction there are alternating ferromagnetic (antiferromagnetic) lines where ferromagnetic (antiferromagnetic) vertical bonds live (we consider that the absolute values of the couplings of all bonds are equal). In this way, the product of bonds on a single plaquette is always $-1$ (three ferromagnetic bonds per plaquette) and therefore the ground-state of the system cannot be just the ferromagnetic one. In fact, by mapping each "unsatisfied" bond to a dimer living on the dual sublattice, we find that each of the degenerate ground-state configuration maps to a hard-core dimer configuration on the square lattice (see Fig.\ref{fig:ffsi}) \cite{liebmann_book}.  

A typical [1n] configuration in the dimer language we used, as clearly depicted in Fig. \ref{fig:ffsi} for one of the four equivalent dimer structures, under $\pi/2$ rotations and sublattice shifts, can be seen in the FFSI picture as ferromagnetic stripes of length $4n$ in the one direction and infinite in the other, separated by antiferromagnetic domain walls. In this way, these ordered states clearly resemble the modulated phases of the ANNNI model in two dimensions. The other three equivalent dimer structures map again to periodic domains in the Ising model, but with more complicated metamagnetic structure. The reason for this seemingly large complexity has to do with the fact that the possible equilibrium configurations have to satisfy the FFSI constraint.

As far as the interactions are concerned, we have the following correspondences: The three-dimer resonance term we used in our construction corresponds to a two neighboring-spins flip process which should, however, respect the FFSI constraint. We should note that the usual single-plaquette resonance move maps to the single-spin flip which is the same as the Ising transverse field usually considered. The $a$ and $b$-terms correspond to domain-wall energies in the FFSI. Both of them have an one-to-one mapping to three-spin interactions but these interactions are also anisotropic (they depend on the distribution of the Ising bonds which we described). The additional interactions that we added to the system, so that to extensively study it, correspond clearly to complicated multi-spin interactions.

\section{Discussion}
\label{sec:discuss}

There are reasons to be optimistic that these 
ideas apply more generally.  For example, as mentioned earlier, we expect that with 
suitable modification of Eq.~\ref{eq:H0}, a similar phase diagram 
may be obtained for a wide variety of off-diagonal resonance terms, 
including the familiar two-dimer resonance of Rokhsar-Kivelson.  This 
is because the perturbation theory is structured so that at any 
order, most of the nontrivial resonances amount to self-energy 
corrections and the resonances driving the transitions are 
comparatively simpler.  The three-dimer resonance of Eq.~\ref{eq:V} is 
analytically convenient as its action is confined to the 
domain wall boundaries.  The two-dimer resonance would involve more 
complicated bookkeeping since we also need to account for 
internal fluctuations of the columnar regions.  

Another reason to expect these ideas to hold more generally is the 
qualitative similarity of this approach to the field theoretic 
arguments in Refs.~\onlinecite{vbs04} and \onlinecite{fhmos04}.  In 
those studies, the following action, in the notation of 
Ref.~\onlinecite{fhmos04}, was used to describe the tilting 
transition in the Rokhsar-Kivelson quantum 
dimer model on the honeycomb lattice (the square lattice is similar 
but with some added subtleties -- see Ref.~\onlinecite{fhmos04}):
\bea
S&=&\frac{1}{2}(\partial_{\tau}h)^{2}+\frac{1}{2}\rho_{2}(\nabla 
h)^{2}+\frac{1}{2}\rho_{4}(\nabla^{2}h)^{2}\nonumber\\&+& g_{3}(\partial_{x} h)
(\frac{1}{2}\partial_{x}h-\frac{\sqrt{3}}{2}\partial_{y}h)(\frac{1}{2}\partial_{x}h+\frac{\sqrt{3}}{2}
\partial_{y}h)\nonumber\\&+& g_{4}|\nabla h\cdot \nabla h|^{2} + \ldots
\label{eq:action}
\eea
where ``$\ldots$'' includes terms that are irrelevant to the present 
discussion (though maybe not strictly ``irrelevant'' in the RG sense).  
In this expression, $h$ is a coarse grained version of the height 
field (Fig.~\ref{fig:height}) and the first line of Eq.~\ref{eq:action}
describes the tilting transition at the RK point\cite{action}, which 
corresponds to $\rho_{2}=g_{3}=g_{4}=0$.  If $g_{3}<0$, the system favors tilted 
states and is similar to our parameter $a-b$.  However, $g_{4}$ prevents the 
tilt from taking its maximal value and in this sense, is similar to 
our terms $c$ and $d$.  The existence of the devil's staircase in 
Ref.~\onlinecite{fhmos04} was established by tuning $g_{3}$ and 
$g_{4}$ so as to stabilize an intermediate tilt and then to study the 
fluctuations about this state.  The staircase arose from a 
competition between these quantum fluctuations, analogous to our 
term $t$, and lattice interactions, (roughly) analogous to our
terms $c$, $d$, $p$, and $q$.  

Another sense in which our calculation is similar to Ref.~\onlinecite{fhmos04} may
be seen by heuristically considering the effect of doping the model.  In particular, consider 
replacing one of the dimers in a staggered strip with two monomers.  If we then 
separate the monomers in the direction parallel to the stripe, a string of columnar bonds will
be created.  If the staggered and columnar bonds were degenerate, then this would 
cost no energy in addition to the cost of creating the monomers in the first place so the 
monomers would be deconfined.  However, in the [1n] phase, the staggered bonds are 
slightly favored so the energy cost $E$ of separating the monomers by a distance $R$ would be 
$E \sim R t^{2n}$.  Hence, the commensurate phases seen in our model are confining with
a confinement length that becomes arbitrarily large for the high-order structures that appear very close to the columnar phase boundary.  This is qualitatively similar to the ``Cantor deconfinement" scenario
proposed in Ref.~\onlinecite{fhmos04}.

However, there are ways in which our calculation is qualitatively different from the above.  Our calculation takes place in the limit of 
``strong-coupling'' where $t$ is
small compared with other terms but 
influences the phase diagram nonetheless because the stronger terms 
are competing.  In contrast, the RK point of a quantum dimer model, by 
definition, occurs in a regime of parameter space where quantum fluctuations 
are comparable in strength to the interactions.  The field theoretic 
prediction requires $g_{3}$ and $g_{4}$ being nonzero so does not 
literally apply at the RK point either but, by self-consistency, 
should apply somewhat ``near'' it.  We may speculate that the tilted states 
being predicted by the field theory are 
large $t$ continuations of states that emerge in the strong coupling 
limit far from the RK point.  However, we reemphasize that our 
calculation is reliable only in the limit of small $t$ and we can not 
be certain which (if any) of our striped phases survive at larger 
$t$.  Another issue is that the phase diagram near the RK point depends strongly
on the lattice geometry and the prediction of a devil's staircase in 
Ref.~\onlinecite{fhmos04} is for bipartite lattices.  In contrast, lattice symmetry 
does not play an obvious role in the present work and it is likely that these ideas
can be made to apply on more general lattices. 

It is also likely that this calculation can be made to work in the 
strong-coupling limits of other frustrated models, for example vertex 
models\cite{ardonne04} and even in higher dimensions.  For example, 
mappings similar to those discussed in Ref.~\onlinecite{rms05} may be used to 
construct an SU(2)-invariant spin model on a decorated lattice that 
displays the same phases.  A more interesting direction would be to study the 
strong-coupling limits of more physical models, for example the Emery model 
of high $T_{c}$ superconductivity\cite{emery87} which also shows an affinity for 
nematic ground states\cite{kivfradgeb04}.  It would also be 
interesting to see whether nematicity is essential, i.\ e.\ whether 
other types of phase separation can occur in a purely {\em local} 
model through effective long range interactions that arise, as in the 
present calculation, from the interplay of kinetic energy and frustration.

%\section{Summary and Conclusion}
%\label{sec:conclusion}

\begin{acknowledgments}
The authors would like to thank David Huse, Steve Kivelson, Roderich Moessner, and 
Shivaji Sondhi for many useful discussions.  In particular, we thank Steve Kivelson for the observation regarding monomer confinement.  This work was supported through the
National Science Foundation through the grant NSF DMR 0442537 at the University of Illinois, and through the Department of Energy's Office of Basic Sciences through the grant DEFG02-91ER45439, through the Frederick Seitz Materials Research Laboratory at the University of 
Illinois at Urbana-Champaign (EF). We are also grateful to the Research Board of the University of Illinois at Urbana-Champaign for its support.
\end{acknowledgments}

\appendix

\section{Details of the second order calculation}
\label{app:ca}

In this section, we explicitly work out the details of the second 
order calculation, including the effect of having $c>a$.  We show 
that the second order phase diagram is qualitatively similar to the 
$c=a$ case presented in the main text, provided that $c$ is not too large 
(the precise condition is obtained below).  Therefore, fine-tuning to 
$c=a$ is purely a matter of convenience.     

With reference to Fig.~\ref{fig:singleflip}, we may calculate the 
difference in energy between the excited and initial states for each 
case.  The unperturbed energies $\epsilon_{i}^{in}$, $i\in\{1,2,3\}$, of 
the initial states in Fig.~\ref{fig:singleflip}-1, \ref{fig:singleflip}-2, and 
\ref{fig:singleflip}-3 are given by Eq.~\ref{eq:energy0} and are 
degenerate when $a=b$.  To calculate the unperturbed energies, 
$\epsilon_{i}^{ex}$ of the excited states, 
we need to examine the interaction energies present in the excited 
states which 
are absent in the initial states and vice versa.  These are shown in 
Fig.~\ref{fig:singleflip} by the red and blue arrows.  Using the 
figure, we obtain:
\bea
\epsilon_{1}^{ex}-\epsilon_{1}^{in} &=& (4d-2a)-(-2a-2b)\nonumber\\&=& 4d+2b
\label{eq:2a}
\eea
\bea
\epsilon_{2}^{ex}-\epsilon_{2}^{in}&=&(4d+2c-2a-2a)-(-2a-2b)\nonumber\\
&=& 4d+2b+2c-2a
\label{eq:2b}
\eea
\bea
\epsilon_{3}^{ex}-\epsilon_{3}^{in}&=&(6d+p-2a-b)-(-2a-2b)\nonumber\\&=& 6d+p+b
\label{eq:2c}
\eea
These are the three possible energy denominators which may enter 
Eq.~\ref{eq:pert2}.  We may 
classify a staggered line based on the dimer arrangement to its 
immediate right, corresponding to cases (1), (2), and (3) in 
Fig.~\ref{fig:singleflip}.  Each staggered line contributes 
$-N_{y}\frac{t^{2}}{\epsilon_{i}^{ex}-\epsilon_{i}^{in}}$.  The 
factor $N_{y}$ is because there are $\frac{N_{y}}{2}$ possible 
rightward resonances of each line and also $\frac{N_{y}}{2}$ leftward 
resonances of the line to its right, which enter with the same 
weight.  Putting everything together, we can update 
Eq.~\ref{eq:energy0}:
\bea
E(N_{s1},N_{s2},N_{s3},N_{c})&=& 
-b\frac{N_{y}N_{x}}{2}+(b-a)N_{y}N_{s}\nonumber\\&-&
N_{y}N_{s1}\frac{t^{2}}{4d+2b}\nonumber\\&-& N_{y}N_{s2}\frac{t^{2}}{4d+2b+2c-2a}
\nonumber\\
&-& N_{y}N_{s3}\frac{t^{2}}{6d+p+b}\nonumber\\&+& O(t^{4})
\label{eq:energy21}
\eea
where $N_{si}$ is the number of staggered lines of type (i) and 
$N_{s}=N_{s1}+N_{s2}+N_{s3}$ is the total number of staggered lines.  
While the zeroeth order term depends only on the number of staggered 
lines, the $O(t^{2})$ piece depends on their distribution and relative 
orientations.
  
Eq.~\ref{eq:energy21} shows that type (1) staggered lines will be 
stabilized more by the perturbation than type (2) lines.  However, the 
zeroeth order term depends only on the total number of lines and type 
(2) lines permit a denser packing of lines.  We 
temporarily ignore type (3) lines.  Depending on $b-a$, the system will favor either the [11] state, the 
columnar state, or the states which maximize the number of type (1) 
lines, which (at second order) are analogous to the maximally 
staggered configurations in Fig.~\ref{fig:phase0} except staggered 
lines are now separated by two columns.  We denote the latter 
collection of states with the label $A_{2}$ (i.\ e.\ alternating states 
where staggered strips alternate with two columns of columnar dimers).  
To determine these boundaries, we first write down the energies of 
these three states to second order in $t$ using Eq.~\ref{eq:energy21}:
\beq
E_{[11]}=-b\frac{N_{y}N_{x}}{2}-\frac{N_{y}N_{x}}{2}(b-a-\frac{t^{2}}{4d+2b+2(c-a)})
\eeq
\beq
E_{A_{2}}=-b\frac{N_{y}N_{x}}{2}-\frac{N_{y}N_{x}}{4}(b-a-\frac{t^{2}}{4d+2b})
\eeq
\beq
E_{col}=-b\frac{N_{y}N_{x}}{2}
\eeq
where we have used $N_{s}=\frac{N_{x}}{4}$ for the [11] state, 
$N_{s}=\frac{N_{x}}{6}$ for the $A_{2}$ states, and $N_{s}=0$ for the 
columnar state.  Comparing these expressions, we obtain the following 
boundaries.  The [11] state is favored when:
\beq
    b<a+\frac{t^{2}}{4d+2b}-\frac{3t^{2}}{2(4d+2b)}(1-\frac{4d+2b}{4d+2b+2(c-a)})
\eeq

The $A_{2}$ states are favored when:
\bea
    a+\frac{t^{2}}{4d+2b}-\frac{3t^{2}}{2(4d+2b)}(1-\frac{4d+2b}{4d+2b+2(c-a)})
    < b \nonumber\\ < a+\frac{t^{2}}{4d+2b} \nonumber\\
\eea
and the columnar state is favored when:
\beq
    b>a+\frac{t^{2}}{4d+2b} 
\eeq
If we require that $(c-a)$ is small compared to $2d+b$ (recall that 
$d\geq a$ by assumption), then we see that the 
region where the $A_{2}$ state is preferred is a small region within 
the phase boundary of Fig.~\ref{fig:phase2}.  The width of this 
region tends to zero as $c\rightarrow a$.  A cartoon of this 
case is given in Fig.~\ref{fig:phase2b}.  On the 
$A_{2}$-columnar boundary, the $A_{2}$ states are degenerate with the 
columnar state and any intermediate state where consecutive staggered lines are 
separated by at least two 
columns.  Similarly, on the [11]-$A_{2}$ boundary, the [11] states are 
degenerate with the $A_{2}$ states and any intermediate states where 
staggered lines of different (same) orientation are separated by two 
(either one or two) columns.    

The collection of 
$A_{2}$ states are degenerate to order $t^{2}$.  As the degeneracy of the 
$A_{2}$ states will be partially lifted 
at fourth order in the perturbation theory, the $c>a$ case
adds complexity without changing the phase diagram qualitatively.
Therefore, purely for convenience, we assume $c=a$ throughout 
the main text.  

As $c$ is made larger, the width of the 
$A_{2}$ region increases but when $c> d+a+\frac{p-b}{2}$, then 
according to Eq.~\ref{eq:energy21}, 
we need to consider type (3) lines.  In this 
case, the tilted state will no longer be favored and, in fact, all of 
the steps of the staircase will lie in the zero winding number 
sector. 

\section{Cancellation of disconnected resonances}
\label{app:disconnect}

We now demonstrate the cancellation of disconnected terms that 
appear at fourth order in the perturbation theory 
(Eq.~\ref{eq:pert4}).  Because our Hamiltonian is local, there are 
linked cluster theorems which ensure that this cancellation occurs 
at any order in the perturbation theory so the contribution to the 
energy is always extensive as it should be.  

Regarding the situation of 
Fig.~\ref{fig:disconnect}, we may write down all fourth order terms 
involving the excited states which we have labelled 1 and 2.  The 
energy numerator will contribute a number of terms, each having energy:
\bea
E &=&-t^{4}\frac{V_{nm}V_{ml}V_{lk}V_{kn}}
{(\epsilon_{m}-\epsilon_{n})(\epsilon_{l}-\epsilon_{n})(\epsilon_{k}-\epsilon_{n})}
\nonumber\\&=&
\frac{-t^{4}}{(4d+2b)^{2}(2(4d+2b))}
\eea
which follows because the energy of the excited states (relative to 
the initial state) are 
$\epsilon_{1}-\epsilon_{0}=\epsilon_{2}-\epsilon_{0}=4d+2b$ and 
$\epsilon_{12}-\epsilon_{0}=2(4d+2b)$.  That 
$\epsilon_{12}-\epsilon_{0}=2(\epsilon_{1}-\epsilon_{0})$ is precisely 
because clusters 1 and 2 are disconnected.  This would not be the 
case if we had a long range interaction.  From 
Fig.~\ref{fig:disconnect}, we readily see that there will be four such 
terms since there are four ways to connect the initial state (0) to 
the excited state (12) and then back to itself.  Therefore, the full 
potentially pathological contribution of the energy numerator is:
\beq
E_{num}=-4\frac{t^{4}}{(4d+2b)^{2}(2(4d+2b))}=-2\frac{t^{4}}{(4d+2b)^{3}}
\eeq
It is evident that there are of order $N^{2}$ of these terms because 
the choice of clusters 1 and 2 were arbitrary.  
The energy denominator (wavefunction normalization) will contribute 
terms, each having energy:
\beq
E=t^{4}\frac{V_{nm}V_{mn}V_{nl}V_{ln}}{(\epsilon_{m}-\epsilon_{n})^{2}
(\epsilon_{l}-\epsilon_{n})}=\frac{t^{4}}{(4d+2b)^{3}}
\eeq
and only resonances from the initial state to excited states 1 and 2 
are involved.  Because the case where $m=l$ gives an extensive contribution to the 
energy, we will be concerned with the case where 
$m$ and $l$ in the above equation are different.  There are two such 
terms because either $m$ or $l$ can correspond to excited state 1 and 
the other to state 2.  Therefore, the full potentially pathological 
contribution of the energy denominator is:
\beq
E_{den}=2\frac{t^{4}}{(4d+2b)^{3}}
\eeq
We see explicitly that the non-extensive contributions of $E_{num}$ and $E_{den}$
precisely cancel.  Clearly this will be case for any choice of 
disconnected clusters 1 and 2 so we have shown that, to fourth order 
in the perturbation theory, the energy correction is extensive, as it 
physically should be.

\bibliography{citpaper.bib}

\begin{thebibliography}{63}
\bibitem[{til()}]{tilt}
\bibinfo{howpublished}{The RK point has the additional feature that the
  degenerate wavefunctions have no local order. At the RK point, the lowest
  energy state in each topological sector is one of the (degenerate) ground
  states of the system. In a given sector, the ground state wavefunction is the
  equal amplitude superposition of dimer coverings in that sector. Such a
  wavefunction will generally give zero expectation value for any local
  operator (except for special cases like the staggered state
  (Fig.~\ref{fig:height}b) which is the only dimer covering in its sector)}.
\bibitem[{foo({\natexlab{a}})}]{footnote:incomm}
\bibinfo{howpublished}{This means that the period of the repeating unit becomes
  infinite. This becomes the case, for example, close to the boundary with the
  columnar phase but may also occur between the other $[1n]$ phases.}
\bibitem[{foo({\natexlab{b}})}]{footnote:14}
\bibinfo{howpublished}{Although we will only ``know'' this at eighth order when
  the degeneracy of the $A_{4}$ states, where adjacent lines are always
  separated by four columnar strips, is lifted in favor the one with maximal
  tilt.}
\end{thebibliography}

\end{document}